\documentclass[9.5pt,journal,compsoc]{IEEEtran}

\pdfoutput=1

\usepackage{amsmath}
\usepackage{amsthm}
\usepackage{amssymb}
\usepackage{graphicx}
\usepackage{subfigure}
\usepackage{color}
\usepackage{balance}
\usepackage{cite}
\usepackage{url}
\usepackage{multirow}

\newtheorem{definition}{Definition}

\newtheorem{lemma}{Lemma}

\newtheorem{theorem}{Theorem}
\usepackage[pdftex,dvipsnames]{xcolor}
\usepackage{xargs} 
\usepackage[colorinlistoftodos,prependcaption,textsize=tiny]{todonotes}
\newcommandx{\unsure}[2][1=]{\todo[linecolor=red,backgroundcolor=red!25,bordercolor=red,#1]{#2}}
\newcommandx{\change}[2][1=]{\todo[linecolor=blue,backgroundcolor=blue!25,bordercolor=blue,#1]{#2}}
\newcommandx{\info}[2][1=]{\todo[linecolor=OliveGreen,backgroundcolor=OliveGreen!25,bordercolor=OliveGreen,#1]{#2}}
\newcommandx{\improvement}[2][1=]{\todo[linecolor=Plum,backgroundcolor=Plum!25,bordercolor=Plum,#1]{#2}}
\newcommandx{\thiswillnotshow}[2][1=]{\todo[disable,#1]{#2}}

\begin{document}
\title{The Tiny-Tasks Granularity Trade-Off \\ {\huge Balancing overhead vs. performance in parallel systems}}

\author{\IEEEauthorblockN{Stefan Bora\thanks{This manuscript is a revised and extended version of~\cite{fidler:tinytasks} which appeared in the IEEE Infocom 2020 proceedings. This work was supported in part by the German Research Council (DFG) under Grant VaMoS (FI 1236/7-1).} \ \  Brenton Walker \ \  Markus Fidler}\\
\IEEEauthorblockA{\textit{Institute of Communications Technology, Leibniz Universit\"at Hannover, Germany} \\
\{stefan.bora, brenton.walker, markus.fidler\}@ikt.uni-hannover.de
}}
\maketitle

\begin{abstract}
Models of parallel processing systems typically assume that one has $l$ workers and jobs are split into an equal number of $k=l$ tasks.  Splitting jobs into $k > l$ smaller tasks, i.e. using ``tiny tasks'', can yield performance and stability improvements because it reduces the variance in the amount of work assigned to each worker, but as $k$ increases, the overhead involved in scheduling and managing the tasks begins to overtake the performance benefit.
We perform extensive experiments on the effects of task granularity on an Apache Spark cluster, and based on these, developed a four-parameter model for task and job overhead that, in simulation, produces sojourn time distributions that match those of the real system.
We also present analytical results which illustrate how using tiny tasks improves the stability region of split-merge systems, and analytical bounds on the sojourn and waiting time distributions of both split-merge and single-queue fork-join systems with tiny tasks.  Finally we combine the overhead model with the analytical models to produce an analytical approximation to the sojourn and waiting time distributions of systems with tiny tasks which include overhead.  Though no longer strict analytical bounds, these approximations matched the Spark experimental results very well in both the split-merge and fork-join cases.  
\end{abstract}

%
%------------------------------------------------------------------------
%------------------------------------------------------------------------
%------------------------------------------------------------------------
%
\section{Introduction}
\label{sec:introduction}

Parallel processing systems improve performance by dividing large jobs into smaller tasks, and distributing those tasks to a cluster of many workers.  To a first approximation, the total amount of processing time does not change, but the amount of time a user must wait for the result can be reduced by orders of magnitude.  Because of the distributed computation, the size of the data set that can be operated on, and held in RAM, is correspondingly increased.  

The first impulse, both in modeling such systems and in practice, is to divide each job into tasks so that it fits evenly on the available workers.  If $k$ is the number of tasks, and $l$ is the number of workers, this means taking $k=l$.  
%Indeed, taking $k<l$ would mean potentially leaving some workers idle, which is not very appealing.  
Using a finer granularity, taking $k>l$, so-called ``tiny tasks'', actually can have a great and positive impact on system performance.  This has been noted by practitioners~\cite{spark:tuning,cloudera:spark-tuning}, but so far only~\cite{fidler:tinytasks}, which this paper is an extension of, provides analytical results relating task granularity to parallel system performance.  
%We will show experimental and simulation results that demonstrate the benefits and trade-offs of using tiny tasks, and present analytical models that provide a deeper understanding of why performance improves.  We develop a model for system overhead based on measurements from a real Apache Spark system, and integrate this into our simulations and analytical models, to better understand what the limits of task granularity are.

\begin{figure}[bt!]
    \centering
    \includegraphics[width=0.9\linewidth]{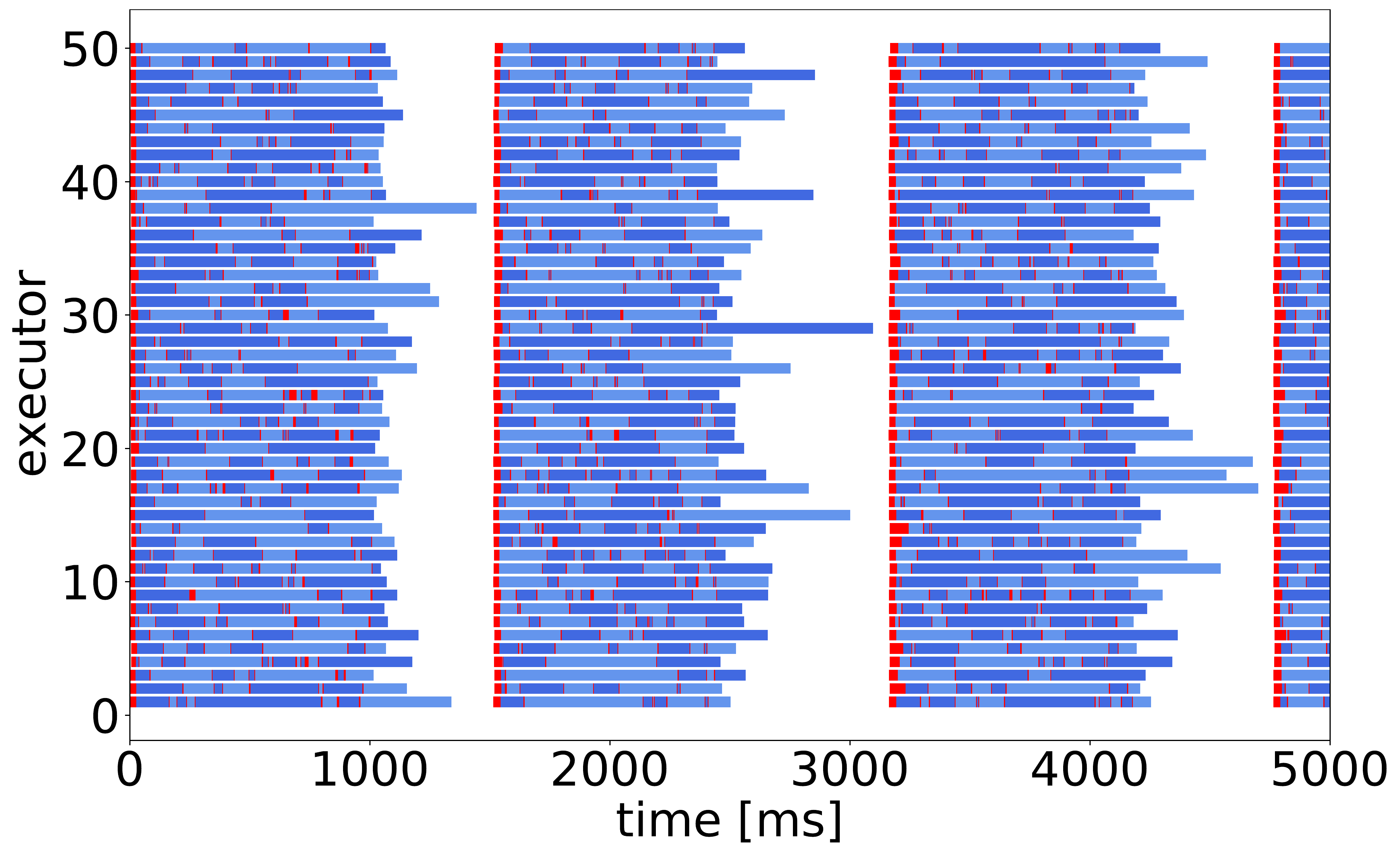}
    \caption{Five second sample of 4 jobs containing 400 tasks per job running on an Apache Spark cluster with 50 executors.}
    \label{fig:tiny_vs_big_tasks_split_merge_big_tasks}
\end{figure}

\begin{figure}[bt!]
    \centering
    \includegraphics[width=0.9\linewidth]{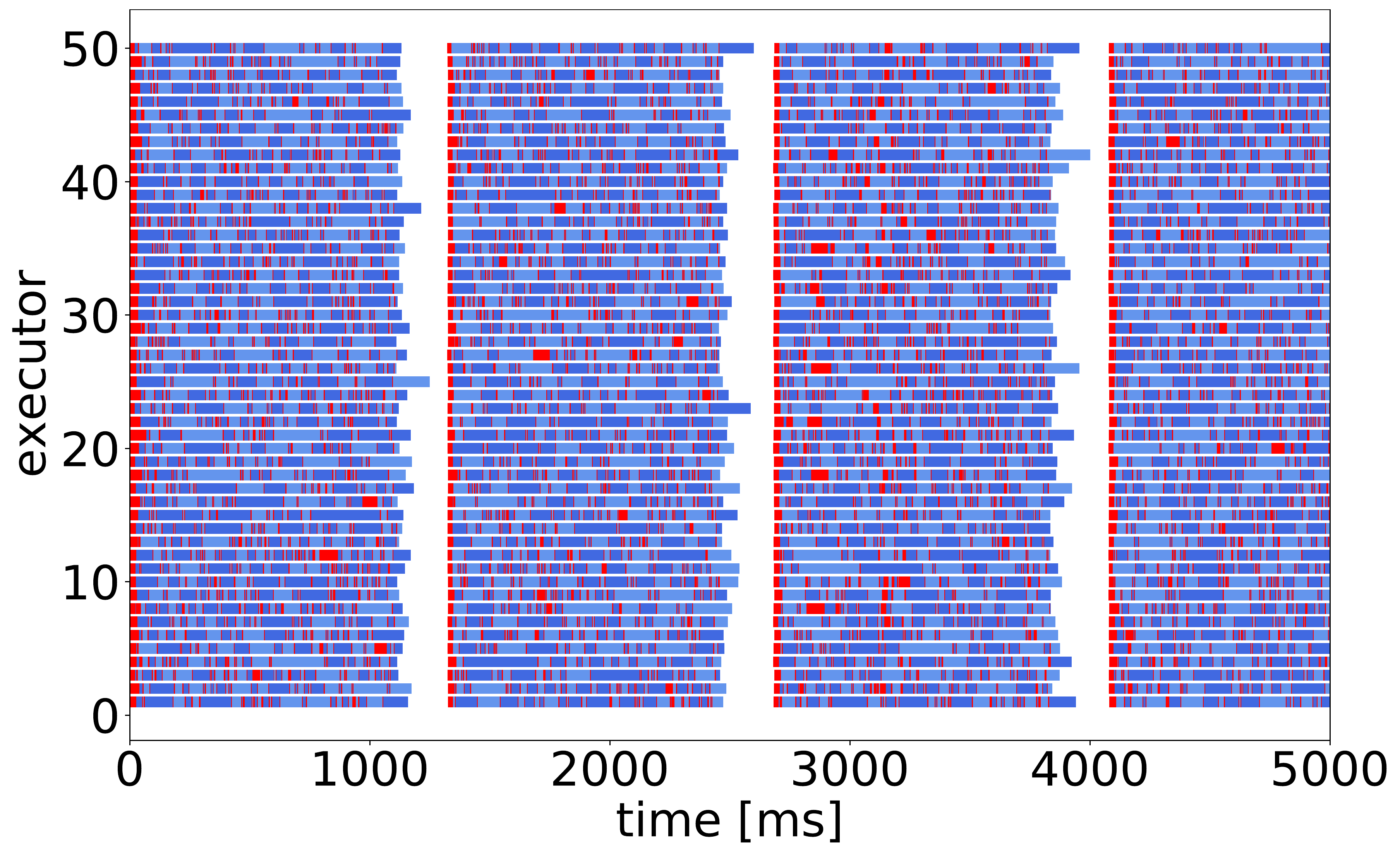}
    \caption{Five second sample of 4 jobs, of the same mean size, containing 1500 tasks per job running on the same cluster.}
    \label{fig:tiny_vs_big_tasks_split_merge_tiny_tasks}
\end{figure}

Figures~\ref{fig:tiny_vs_big_tasks_split_merge_big_tasks} and ~\ref{fig:tiny_vs_big_tasks_split_merge_tiny_tasks} show diagrams of the activity of 50 executors (workers) in a standalone Apache Spark cluster servicing a sequence of four jobs, divided into tasks with different levels of granularity.  In this case the jobs are processed as if they are being submitted from a single-threaded driver program, so each job does not begin until the previous one departs.  In Fig.~\ref{fig:tiny_vs_big_tasks_split_merge_big_tasks} the jobs are divided into 400~tasks, and in Fig.~\ref{fig:tiny_vs_big_tasks_split_merge_tiny_tasks} they are divided into 1500~tasks.  It is immediately apparent that more executors spend much more time idling in the case with coarser task division.  Further, in the case with finer task division, the fourth job is almost complete after 5000~ms, whereas in the coarser case, the fourth job is just starting service.

The primary reason this happens is that when smaller tasks are used, the variance in the amount of the work assigned to each worker decreases.  On the other hand, in real systems, there will always be some trade-off limiting the performance gains. As tasks are made smaller and smaller, at some point the overhead of scheduling and gathering results from tasks will dominate the operation.

\begin{figure}
\centering
\includegraphics[width=0.95\linewidth]{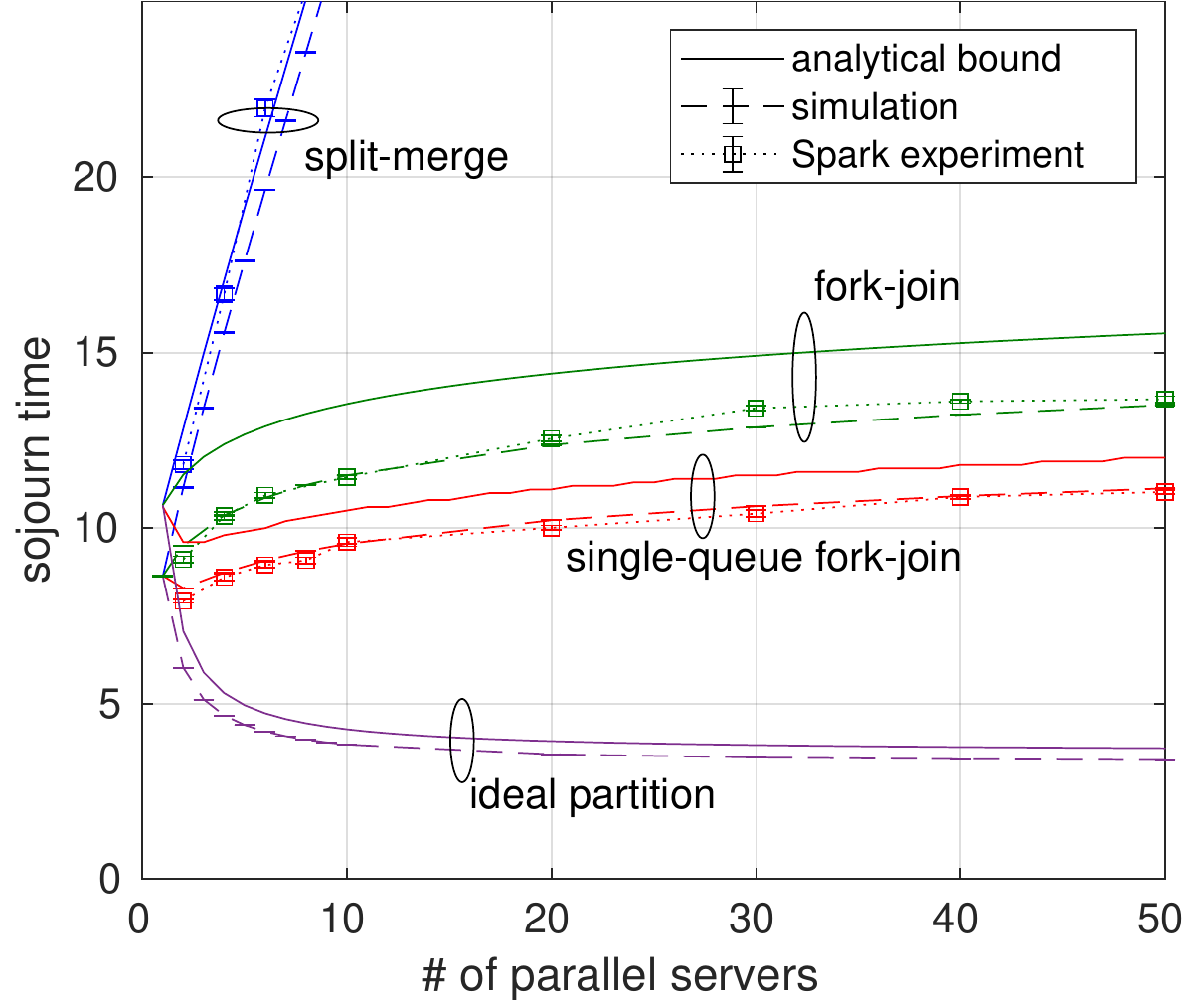}
\caption{Scaling of sojourn time quantiles of the conventional ($k=l$) split-merge, fork-join, and single-queue fork-join models for varying degrees of parallelism.  Exponential arrivals ($\lambda = 0.2$) and task service times ($\mu=1.0$).  
We also include results for the ideal job partition, where a job is partitioned into $l$ equally sized tasks.
%The $10^{-3}$ quantile of sojourn time is plotted. In addition to analytical and simulation results, this shows experimental results from a Spark cluster marked with squares.  We also include results for a parallel processing system with an ideal job partition, where a job is partitioned into $l$ equally sized tasks.
%slightly overestimate the $(1-\varepsilon)$-quantiles obtained from simulation (thin lines). For comparison, we also include results for a parallel processing system with an ideal (hypothetical) job partition, where a job is partitioned into $l$ equally sized tasks. The ideal job partition marks the best case. Note that in case of the ideal partition all three aforementioned models perform identically.
}
\label{fig:sojourntimecomparison}
\end{figure}

This paper explores the issue of task granularity in three domains.  First we perform experiments with real Apache Spark systems to understand the practical limits of the performance gains that can be realized from refined task granularity, and the sources of overhead.  Similar measurements have been done before, either to develop practical guidance for improving cluster performance \cite{spark:tuning,cloudera:spark-tuning}, or in evaluating distributed schedulers that could support larger degrees of parallelism~\cite{ousterhout:sparrow}.  In our case we focus on statistically principled experiments that use tasks with service times drawn from controlled distributions.  Second we use simulation to study the effects of task tinyfication, and model the effects of different types of scheduling overhead.  Finally we derive queuing theoretic results.  In analyzing the split-merge system with tiny tasks, we will formulate the tiny tasks model as a direct refinement of the ``big tasks'' model.  We will derive expressions for the stability region in both cases.  In the limit as $k\rightarrow\infty$, the stability region approaches one.  Then we derive statistical performance bounds for both the split-merge and fork-join models with tiny tasks, and show how their performance improves over the equivalent big tasks model.  As $k\rightarrow\infty$, the performance approaches that of the {\bf ideal job partition}.  This is achieved when the jobs are partitioned into $k=l$ equally sized tasks.  Finally we develop a model for system overhead based on measurements from a real Apache Spark system, and integrate this into our simulations and analytical models, to better understand the limits of task granularity.

%In all cases we consider systems that process jobs from a single-threaded driver program (corresponding to the classical Split-Merge model), and systems that process jobs from a multi-threaded driver (the more complicated Single-Queue Fork-Join model). 

%
%------------------------------------------------------------------------
%------------------------------------------------------------------------
%

\subsection{Systems, models, and stability regions}
\label{sec:systems-models-stability}

Our experiments with real systems are focused on Apache Spark.  Spark is a popular parallel processing engine that implements a map-reduce API~\cite{spark-usenix,spark-url}.  Fig.~\ref{fig:sojourntimecomparison} shows that, depending on the constraints put on the system, a Spark program may exhibit the scaling behavior of split-merge, fork-join, or single-queue fork-join, three different models of parallel systems with quite different scaling behavior.  We will summarize those models and some prior analytical work here.

Across all of the models we will discuss there are some common random processes, the arrival, departure, and task service process.  
Let $A(n)$ for $n \ge 1$ denote the {\bf arrival time} of job $n$, and $D(n)$ the {\bf departure time}.  One key performance statistic we will focus on is the {\bf job sojourn time}, $T(n)=D(n)-A(n)$.
Given $l$ parallel servers and $k$ tasks per job, let $Q_i(n)$ denote the {\bf task service time} of task $i \in [1,k]$ of job $n \ge 1$.  
The {\bf workload} of job $n$, $L(n)=\sum_{i=1}^{k}Q_i(n)$, is defined to be the total of the service required by all of its tasks.  The {\bf job service time} $\Delta(n)$ is the total time a job spends in service.  That is, the time between when its first task begins service and when all of its tasks finish service.  Note that for the parallel models, $L(n)$ and $\Delta(n)$ are not necessarily equal.

\begin{figure}
    \centering
    \subfigure[Fork-join model.]{
        \includegraphics[width=0.9\linewidth]{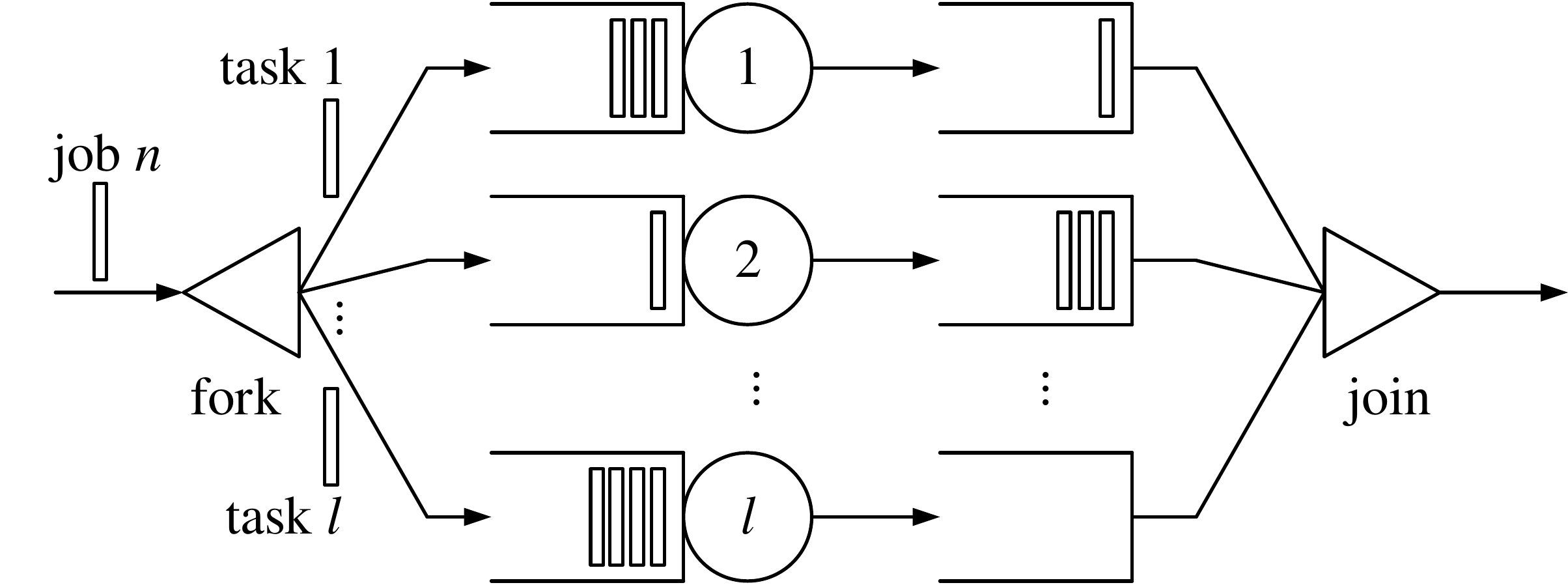}
        \label{fig:forkjoinsystem}
    }
    % \hspace{50pt}
    \subfigure[Split-merge model.]{
        \includegraphics[width=0.9\linewidth]{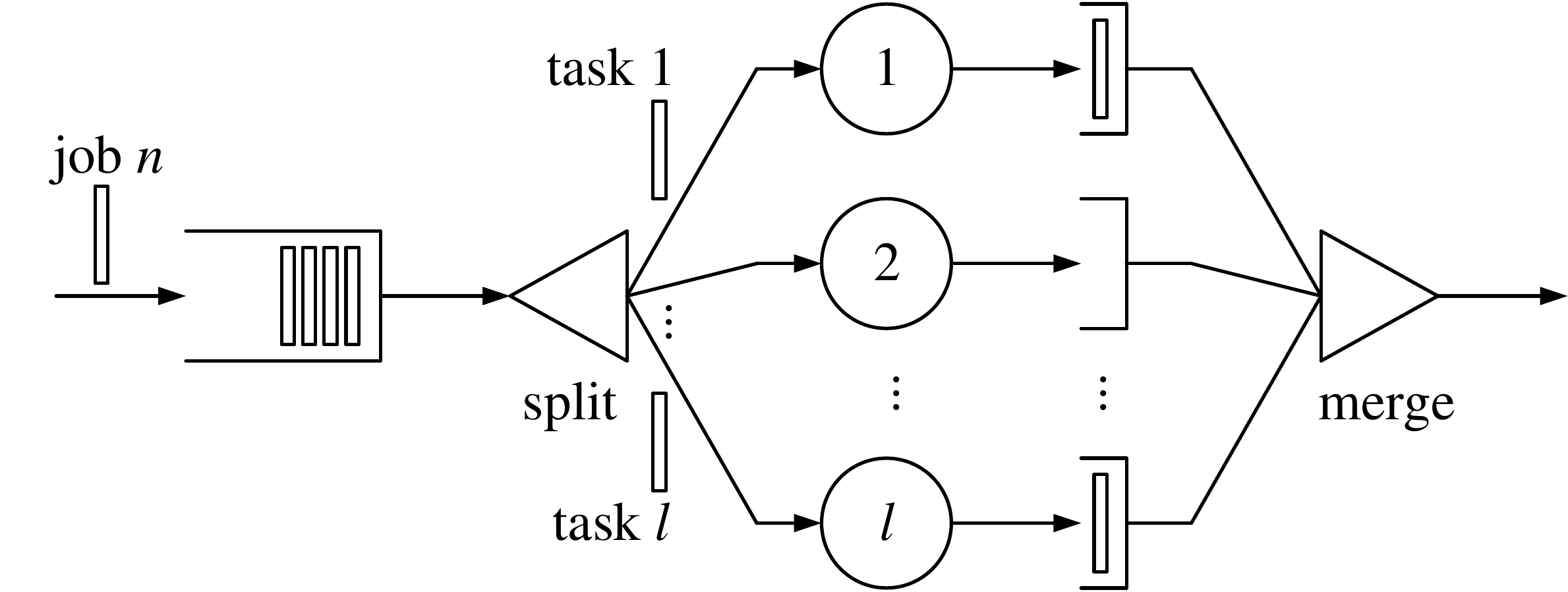}
        \label{fig:splitmergesystem}
    }
    \caption{Models of parallel systems.}
    \label{fig:forkjoin-splitmergesystem}
\end{figure}

%The {\bf workload} of a job $L(n)=\sum_{i=1}^{k}Q_i(n)$ is defined to be the total of the service required by all of its tasks.  The {\bf job service time} $\Delta(n)$ is the total time a job spends in service.  That is, the time between when its first task begins service and when all of its tasks finish service.  Note that for the parallel models, $L(n)$ and $\Delta(n)$ are not necessarily equal and $S(m,n)$ may not generally be defined in increments of $\Delta(n)$.

Fig.~\ref{fig:forkjoinsystem} shows a schematic of the {\bf fork-join} model.  Jobs enter the system and are divided into $k$ tasks (fork) that are assigned one by one to $l=k$ servers.  Once all $k$ tasks of a job are serviced, the job leaves the system (join).  The difficulty in analyzing fork-join systems arises from the synchronization constraint of the join operation, and an exact solution is only known for the M$\mid$M$\mid$1 case with  $k=l=2$~\cite{flatto:forkjoin,nelson:forkjoin}.  For broader classes of systems, a variety of approximation techniques have been used~\cite{lebrecht:forkjoin,ko:forkjoin,tan:forkjoin,varma:forkjoin,varki:forkjoin,alomari:forkjoin}.  More recently several researchers have used stochastic network calculus to derive performance bounds~\cite{kesidis:forkjoin,rizk:forkjoin,fidler:multiserver,fidler:tinytasks,walker:bem}.

A schematic of the split-merge model is shown in Fig.~\ref{fig:splitmergesystem}.  The {\bf split-merge} model, also referred to as ``blocking fork-join'' in~\cite{rizk:forkjoin}, has the additional synchronization constraint that the system is blocked until the current job departs.
%Hence a new job starts service only when all servers are idle.  
Parallel systems that behave like split-merge arise easily in practice.  For example any Spark program with a single-threaded driver program, or a single user submitting jobs from a data analysis notebook.  Related, but not exactly the same, many machine learning algorithms require that all tasks in a job start and depart simultaneously to facilitate communication and data transfers between the workers~\cite{horovod-arxiv,spark:horovod-uber,Foldi-2020}.

The analysis of the split-merge model turns out to be much simpler because it behaves like a single-server system with service times given by the service time of the largest task of each job~\cite{harrison:splitmerge,joshi:knforkjoin,lebrecht:forkjoin}.
The problem with the conventional ($k=l$) split-merge model is that it becomes unstable for utilizations well below one, and it becomes unstable more quickly as the degree of parallelism increases, as seen in Fig.~\ref{fig:sojourntimecomparison}.  This has led some researchers to discount the model as impractical~\cite{rizk:forkjoin}.

A third model arises in practice.  When jobs are submitted by a multi-threaded driver program, MapReduce engines such as Apache Spark and Hadoop MapReduce behave like a {\bf single-queue fork-join} system, where all tasks are held in a single FIFO queue and assigned to servers as they become available~\cite{walker:icfc2017}.  Compared to the fork-join model, where tasks are bound to particular servers and a large task can block tasks of subsequent jobs, in the single-queue fork-join model small jobs can overtake jobs with large straggler tasks.  Mean sojourn times for such systems are derived in~\cite{nelson:parallelprocessing}, and bounds on the sojourn time are derived using network calculus in~\cite{fidler:multiserver}.
%The single-queue fork-join model is non-idling in the sense that no server will idle if there are unfinished tasks.  Performance bounds for the $l=k$ case of the single-queue fork-join model were derived in~\cite{fidler:multiserver}.

Fig.~\ref{fig:sojourntimecomparison} shows how job sojourn time scales with the number of servers for these three models in the case with $k=l$ and exponential inter-arrival and task service times.
%This is for the least granular case, where the number of tasks per job, $k$, equals the number of servers, $l$.
The plot shows performance bounds derived using network calculus in~\cite{rizk:forkjoin} and~\cite{fidler:multiserver}, simulation results, and experimental results from an Apache Spark cluster, and demonstrates that a Spark system may behave like any of these three parallel models, depending on how it is configured and how the driver program behaves.
For comparison, the plot includes the equivalent sojourn time statistics for the ideal job partition.  Both fork-join systems show a logarithmic increase in sojourn time as the degree of parallelism increases because of the synchronization constraint~\cite{baccelli:forkjoin, rizk:forkjoin}. The performance of the split-merge system appears catastrophic by comparison.

%
%------------------------------------------------------------------------
%------------------------------------------------------------------------
%

\subsection{Introduction to tiny tasks}

It is no surprise that map-reduce practitioners have devised methods to increase the performance of their systems, even in the split-merge case.  The simplest of these is to partition jobs into a larger number of tasks than there are servers, ${k > l}$.  A common guideline is that the number of tasks should be about three times the number of servers, i.e., $k\approx 3l$~\cite{spark:tuning}, or optimized through trial and error~\cite{cloudera:spark-tuning}.  Some researchers have proposed even more extreme task granularity, $k \gg l$, coining the term ``tiny tasks''~\cite{ousterhout:tiny-tasks}.  

We would like to formalize our conception of the tiny tasks regime.  
%This notion is applicable to the Split-Merge, Fork-Join, and Single-Queue Fork-Join models.
We assume a system with $l$ servers and jobs partitioned into $k \ge l$ tasks, where $k$ may be orders of magnitude larger than $l$.  We define $\kappa = k/l$ to be the {\bf factor of tinyfication} (i.e., $\kappa$ is the average number of tasks from each job served by each of the servers).  
The least granular case, where $\kappa=1$ and $l=k$, gives us the conventional parallel models.  We will refer to the tasks in this case as ``big tasks''.  When $\kappa>1$ we refer to them as ``tiny tasks''.

In the split-merge and fork-join cases, there may be some ambiguity as to how the tiny tasks variants of these models work.  In both cases we model the tiny tasks system to behave in the same way that Spark would behave when given $k>l$ tasks per job.

\begin{figure}
    \centering
    \includegraphics[width=0.9\linewidth]{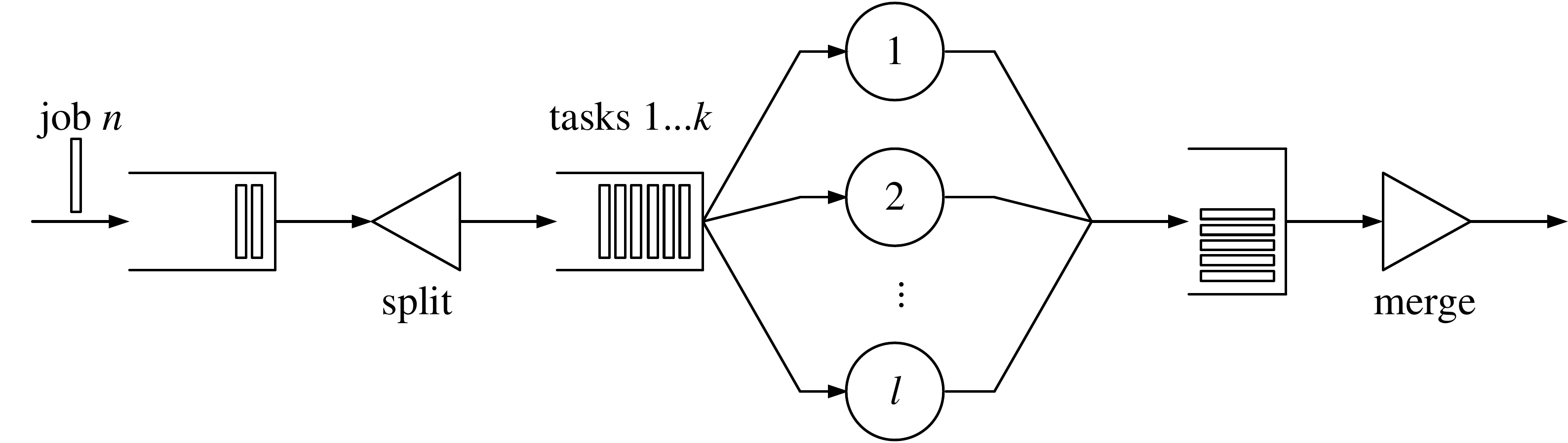}
    \caption{Split-merge model with tiny tasks.}
    \label{fig:tinytaskssystem}
\end{figure}

A schematic of the split-merge tiny tasks model is shown in Fig.~\ref{fig:tinytaskssystem}. Jobs are stored in a job queue, and if there is no job in service, the head-of-line job is partitioned into $k$ tasks (split) which are then stored in the task queue.  Since all servers are idle at the start of a job, the first $l$ tasks start service immediately.  Whenever a server finishes a task, it fetches the head-of-line task from the task queue.  When all $k$ tasks have finished service, the job leaves the system (merge) and the next job, if any, is partitioned and starts service.

In the case of fork-join, using tiny tasks only makes sense in the context of a single-queue model.  In the standard fork-join model, where tasks are bound to specific servers on arrival, tiny tasks would make no difference.  Therefore, throughout this paper, when we refer to fork-join with tiny tasks, it should be understood that we are referring to the single-queue fork-join model.

%
%------------------------------------------------------------------------
%------------------------------------------------------------------------
%------------------------------------------------------------------------
%

\section{Tiny tasks on Apache Spark}
\label{sec:tiny_tasks_on_real_system}

Apache Spark is a popular parallelized data analytic platform implementing the map-reduce paradigm, and therefore well suited to evaluate the performance of tiny tasks on a real cluster system.  Some papers come to the conclusion that the scheduling and bookkeeping overhead required by tiny tasks outweighs the advantages on real systems~\cite{ousterhout:sparrow} or make it impracticable to run on platforms with a centralized scheduler like Apache Spark~\cite{totoni:against-tiny-tasks}.  In this section we will report on our extensive measurements using varying number of tasks per job, investigate the sources and behavior of overhead, and present an overhead model suitable for use in simulation and analytical models, that produces sojourn time distributions matching real experiments.

%
%------------------------------------------------------------------------
%------------------------------------------------------------------------
%

\subsection{Execution model of Apache Spark}
\label{sec:spark_execution_model}

\begin{figure} 
    \centering
    % \subfigure[]{
    \includegraphics[width=0.85\linewidth]{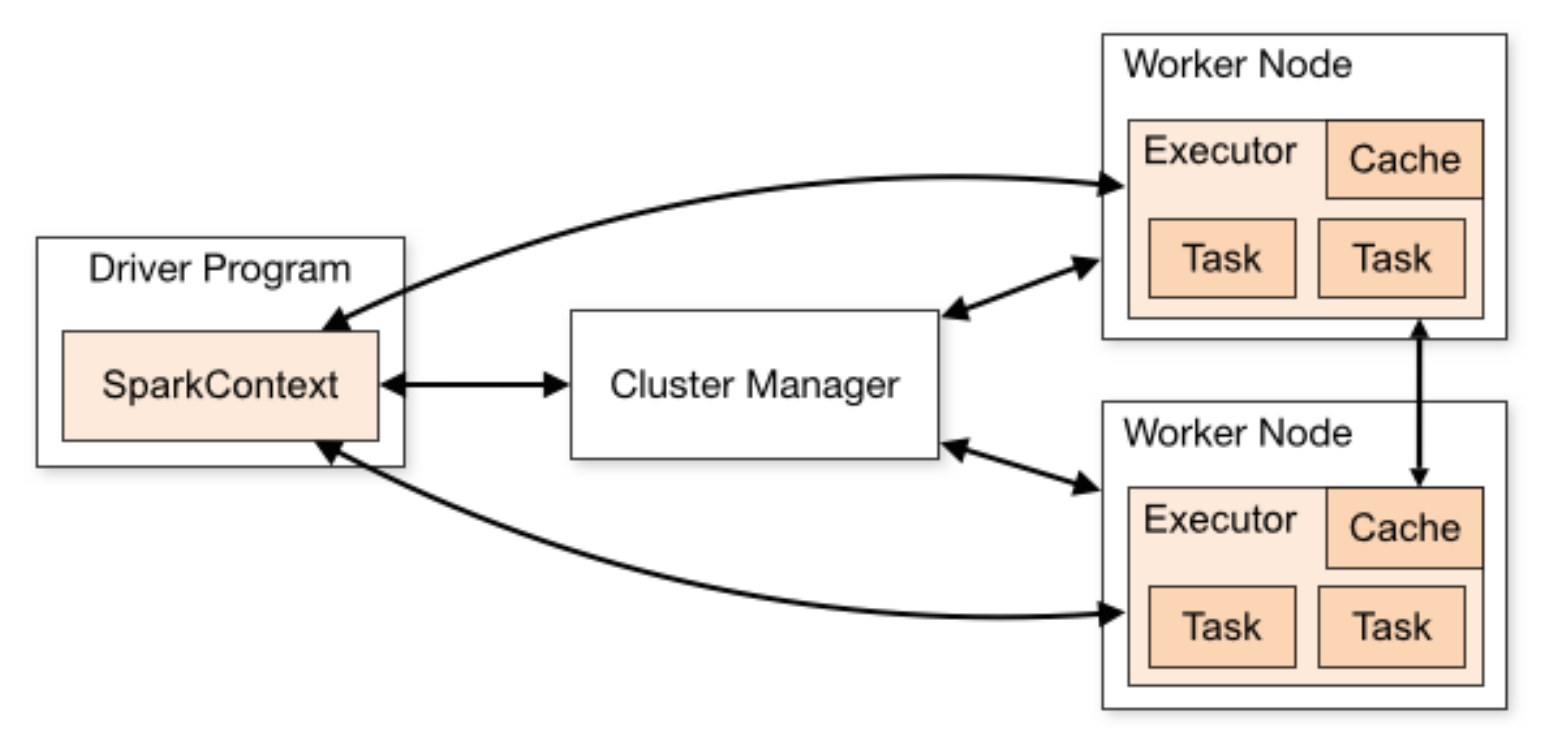}
    % \label{fig:sojourntime_sim_vs_spark_099}
    % }
    \caption{Schematic of the spark model ~\cite{spark:cluster-overview}.}
    \label{fig:spark_model}
\end{figure}

We experiment with Spark in stand-alone mode with the default scheduler~\cite{spark:cluster-overview}.
Fig.~\ref{fig:spark_model} shows a schematic of the Apache Spark components.
A cluster consists of numerous worker nodes which offer their resources to the cluster manager.
The cluster manager allocates the resources to an application running a {\bf driver} program.  Depending on the requested resources, each worker node can host one or more {\bf executors} which connect to the SparkContext in the driver program. %, which submits the actual tasks {\color{red} (not jobs?)}.

% The workers create executor instances which connect to the driver. 
% Depending on the resources of a worker and the requested executor configuration, there can be multiple executors per worker. 
%%%%%%%%%%%%%%%%%%%%%%% This could be part of the experiment environment
% Because these executors run on the same Java Virtual Machine (JVM) of the worker, in our experiments every worker holds only a single executor. To make sure that the workers are independent of each other, even running on the same node, every worker is put in a docker container.

% In our experiments every worker is put in a docker container and can run one executor. This ensures that every executor and it's JVM is isolated from each other even if they are running on the same host.
% As cluster manager we use the standalone manager which comes included into Spark. 
% Running a program on Spark will start a driver application which requests the specified amount of resources from the Spark master. The driver holds a queue with jobs that need to get scheduled on the cluster.
% The driver has a single queue for the jobs which needs to get scheduled on the cluster and splits them into smaller tasks according to the configuration and the program code. The tasks will then get scheduled on free executors in the cluster depending on restrictions of the application like the locality level and the current locality of the data.

The SparkContext in the application maintains a queue of jobs waiting to be scheduled.
Every Spark job operates on a Resilient Distributed Dataset (RDD).  An RDD is a memory representation abstraction~\cite{zaharia:rdds} consisting of multiple partitions which can be distributed throughout the cluster.  There are two types of operations that can be executed on an RDD.  {\bf Transformations} like \texttt{map} are used to create new RDDs from existing ones in a lazy operation.  {\bf Actions} execute the calculations on the cluster and block the thread until the result is returned.  Spark internally divides jobs into one or more stages consisting of tasks which can run independently from each other.  The resulting execution plan is represented as a Directed Acyclic Graph (DAG).  A stage can include multiple transformations and ends with an action.  The most common reason for multiple stages are operations that cause a shuffle, such as \texttt{reduceByKey}, which lead to repartitioning of the RDD.  The DAG ensures that when there is a sequential dependency, tasks of the next stage can only be executed after the previous stage has finished.
% Running actions in a single thread causes the system to behave like a split-merge system.  Executing actions in multiple threads on the other hand generally makes the system behave like the single-queue fork-join model, as tasks of subsequent jobs can be scheduled before the previous job is done. {\color{red} (this keeps getting repeated...)}

%
%------------------------------------------------------------------------
%------------------------------------------------------------------------
%

\subsection{Overhead}
\label{sec:overhead}
%Overhead in distributed systems can be a problem.  
%We consider any time taken by the scheduler or workers to do anything other than execute the tasks to be overhead.  
%
In any cluster with a central scheduler, like Apache Spark, there is overhead which cannot be avoided.  For example a task must be scheduled for execution, and its code and data need to be serialized and sent to an executor.  Depending on the number of tasks and executors handled by the scheduler, this can result in substantial overhead relative to the actual task execution time.  Especially with extremely small tasks with millisecond run times.

Some of the scheduling delays depend on the speed of the worker nodes and the network.  The serialization technique used is also of crucial importance because of the associated processing and transmission time~\cite{kryo}.
After receiving a task, the executor must deserialize it before the actual workload can run.  Depending on the workload, additional data may have to be fetched over the network or loaded from disk.  
%\improvement{Add references to papers which investigated this} 
After the executor services the task, the result has to be serialized and sent to the driver, and written out to a disk or kept in memory.  After a job finishes, the scheduler has to collect the results of its tasks and return a result that depends on the executed action.

% Due to the implementation of Apache Spark there is an additional overhead to fetch jar files depending on the application. Because this is done only once and is highly dependent to the system, it is not considered in this paper. Also loading and moving experimental data is beyond the scope of our experiments.

\begin{figure}
    \centering
    \includegraphics[width=0.950\linewidth]{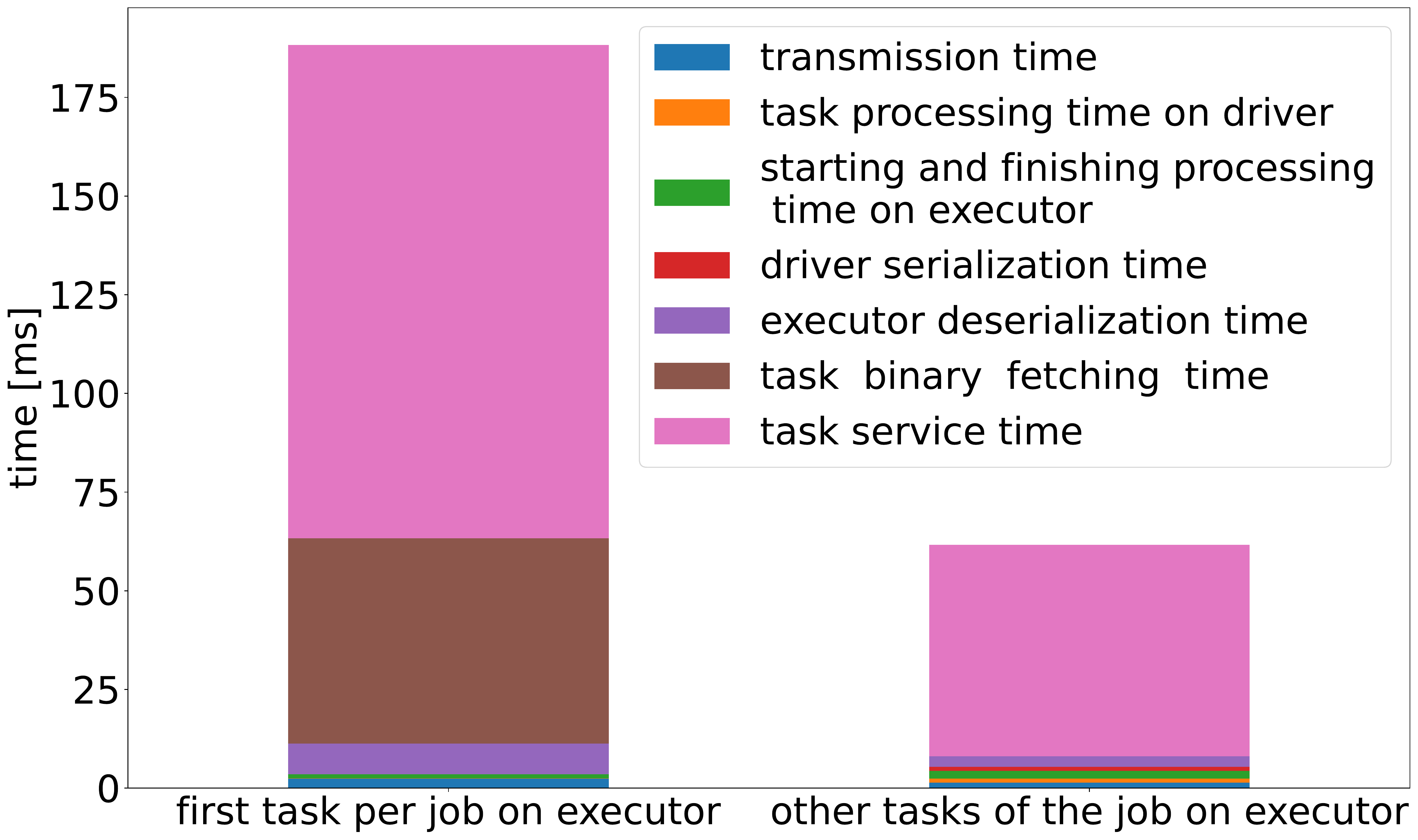}
    \caption{Example of the Spark task duration with 1500 tasks/job with the remaining parameters as in Fig.~\ref{fig:sojourntimeconvergence}. The left bar shows a sample of the first task of a job which ran on an executor whereas the right bar shows an example of a task which is not the first task.}
    \label{fig:detailed_task_overhead}
\end{figure}

There are two classes of overhead we will need to consider.  {\bf Task-service overhead} is attributable to individual tasks and blocks an executor core from servicing the next task.  This type of overhead is fairly well measured and represented in the Spark UI and logs.  {\bf Pre-departure overhead} is the additional delay after all tasks of a job complete, but before a job can depart.  It does not necessarily block the tasks of subsequent jobs from being serviced.  Since it is not attributable to individual tasks during their run time, it is not well represented in the Spark UI and task metrics.  We deduce, its statistical properties based on the overhead model required to fit the experimental sojourn time distributions on real Spark systems.  This will be discussed in more detail in section~\ref{sec:overhead-distribution}.

\begin{figure*}
\centering
\subfigure[Split-merge]{

\includegraphics[width=0.45\linewidth]{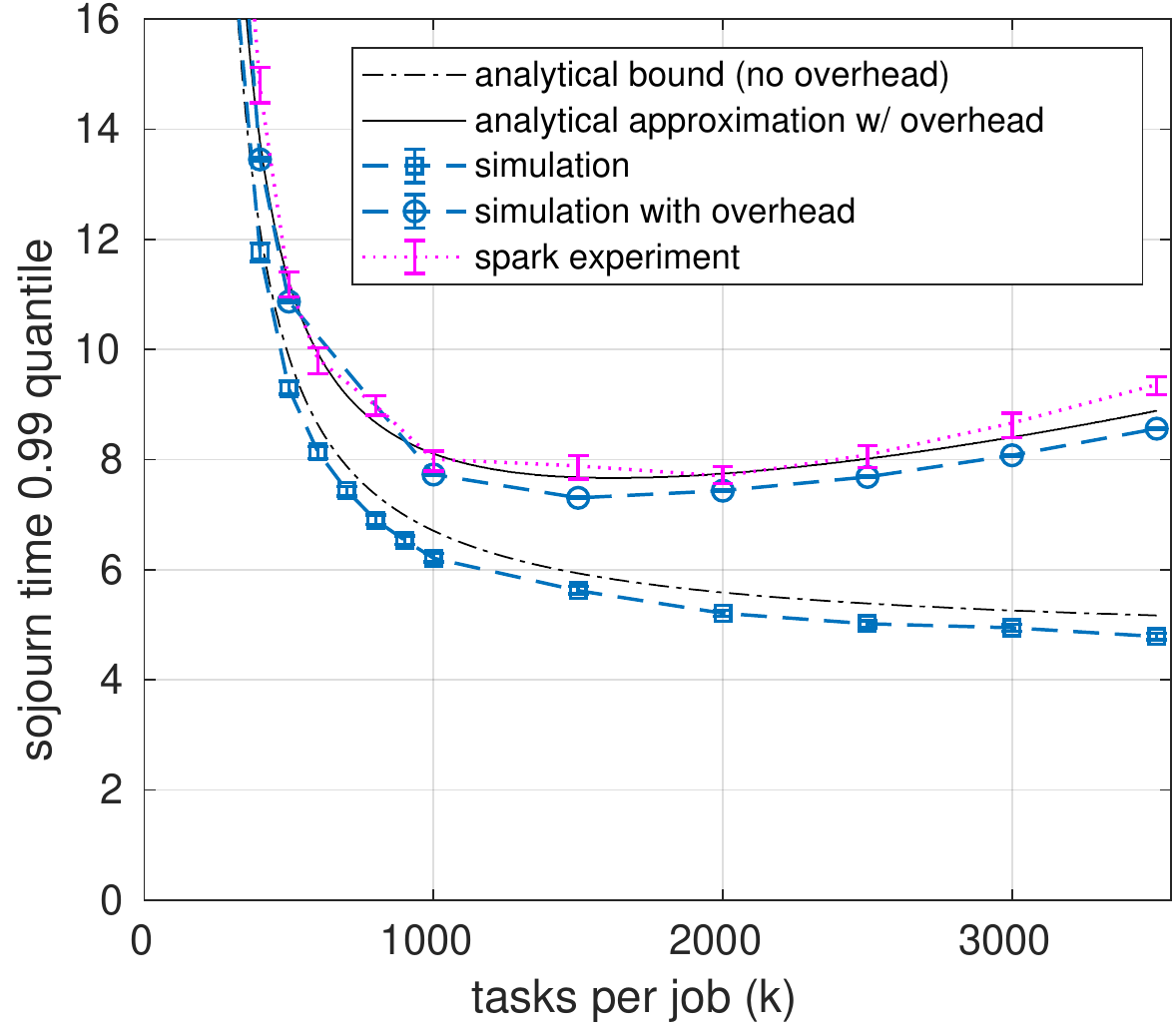}
\label{fig:sojourntime_sim_vs_spark_sm}
}
%\hspace{50pt}
\subfigure[Fork-join]{
\includegraphics[width=0.45\linewidth]{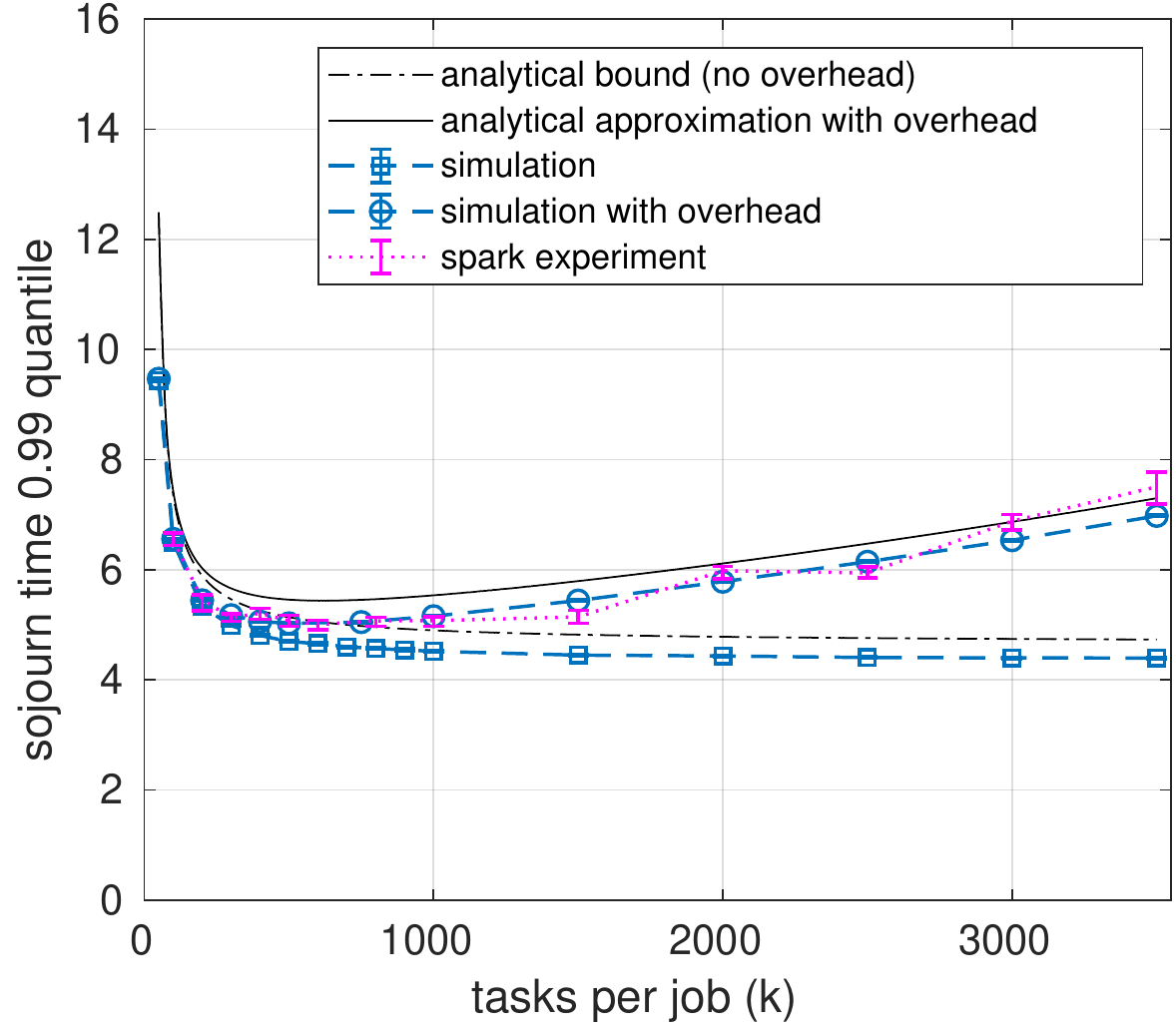}
\label{fig:sojourntime_sim_vs_spark_fj}
}
\caption{Comparison of the sojourn time bounds of the single-queue fork-join and split-merge models with $l=50$ servers and $k$ tiny tasks. Jobs have exponential inter-arrival times with parameter $\lambda=0.5 \text{s}^{-1}$ and are composed of $k$ tasks with exponential service times with parameter $\mu = \frac{k}{l} \text{s}^{-1}$.  The analytical bound and analytical approximation with overhead will be explained in sections~\ref{sec:stabilizing-splitmerge}, \ref{sec:sqfj-tiny-tasks-analysis}, and~\ref{sec:overhead-analyitical-models}.}
\label{fig:sojourntime_sim_vs_spark}
\end{figure*}

Fig.~\ref{fig:detailed_task_overhead} shows a breakdown of the execution and overhead times of two successive tasks running on a Spark cluster.
%As these occupy the cores of the executor and block subsequent tasks from starting, they are all (except the task execution time itself) examples of task-service overhead.  
They can be categorized as follows:
\begin{LaTeXdescription}
\item[Transmission time:]
The time used to send the task information to the executor and to notify the driver about a finished task.  When the task result size is below a hard-coded threshold, it is included in this transmission as well.  If the result is larger, a reference is included and the result is fetched asynchronously later.
\item[Task processing time on driver:] Time between fetching the task out of the queue and sending it to an executor, and the time after receiving the message of a finished task and marking it as finished on the driver.
\item[Starting and finishing processing time on executor:] Time required for housekeeping on the executors. Includes parsing the task message from the driver, decoding the task description and putting the task in the execution queue before running the task. After the task execution is done the executor marks the task as finished, removes it from the list of running tasks and sends the task status including the task result, depending on it's size and the configuration, directly or using the block manager, to the driver.
\item[Driver serialization time:] 
The serialization time of a task on the driver.  In Spark this is done in two parts.
First the task itself is serialized.
%and can contain different content depending on the Task.
%Currently the standard version of Spark offers only ShuffleMapTask and ResultTask which contain the same data except the preferred task execution location and the output id, which are not used in ShuffleMapTask. 
The main content is the task binary, which is a broadcast variable, the RDD (in serialized form only the identifiers), and the accompanying metadata such as stage and task ID.
In a second step the task description is serialized.  This object includes some redundant data such as the task and partition IDs, but also additional data such as the executor ID where the task will run, files and jars which needed to be added, some resources, and the serialized task.
%The reason for the redundancy is that the classpath and the loaded files might need to be modified before the task gets deserialized.

%Serializing the task on the driver. This is done in two parts in Spark. First the task is serialized and then additional task information are serialized. The serialized data includes metadata about the task like the id, needed files, and additional programming code like jar files, and also the serialized task {\color{red} (so besides the jar, needed files, task metadata, what needs to be serialized?)}.

\item[Executor deserialization time:] The time needed to deserialize the task on the executor.  The standard Spark metrics include in this the time to fetch the task binary, even if this happens remotely.  In our statistics we separate this into the item below.
%For a better separation we include this under task binary fetching time.

\item[Task binary fetching time:] Time required to fetch the serialized task function and RDD stored in a broadcast variable.  Each executor only has to fetch this remotely once.  After that, the local copy is used. 

\item[Task execution time:] The time the executor spends actually executing the task.
\end{LaTeXdescription}

The task duration is the sum of the components above.  As a sanity check, this sum should equal the time between the scheduling of the task until the driver receives its result.  We consider everything in the list above, except task execution time, to be overhead.

Next to the listed overhead there are additional types of overhead not considered in our experiments. The time to load RDD data from the disk or over the network is beyond the scope of our investigations.  Also the time to fetch additional jar files is not considered because it appears only once per application.
%and is related to the implementation of Spark.  

For clarity in the subsequent discussion, we need to introduce some terminology and notation for overhead and how it relates to the task service time process.  Recall that the task service time, $Q_i(n)$, is the time between when the scheduler takes up task $i$ of job $n$ and when the worker becomes available to service the next task.  In real systems we consider these service times as comprising two components.
\begin{equation}
    Q_i(n) = E_i(n) + O_i(n)
\end{equation}
where $E_i(n)$ is the {\bf task execution time} of task $i$ of job $n$, and $O_i(n)$ is the {\bf task overhead}.  When doing experiments on a Spark cluster, we can control the execution times of the tasks and draw them from known distributions, but the tasks' overhead is simply a property of the system and we need to measure and model it.

% \textbf{Transmission time:} The time used to send the task information to the executor and notifying driver about a finished task. In our experiments the message to the driver directly included the result of the execution because of its low size.
% \textbf{Task processing time on driver:}
% \textbf{Starting and finishing processing time on executor:}
% \textbf{Driver serialization time:}
% \textbf{Encoding time:}
% \textbf{Only executor deserialization time:}
% \textbf{Getting block data:}
% \textbf{Executor runtime:}

% Maybe put this in the further possible extensions
% Also the data locality problem which can create stagglers in cluster systems when the data is located on one executor while another one is idle is beyond the scope of the investigation if tiny tasks can bring an advantage on current cluster systems.

%
%------------------------------------------------------------------------
%------------------------------------------------------------------------
%

\subsection{Experiment environment}
\label{sec:environment}
For our Spark experiments we allocated 13 nodes from our institute's Emulab testbed~\cite{emulab}, connected via a 1Gbit/s network. 
%Table ~\ref{tab:hardware_configuration} shows an overview of the hardware and software configuration.
To run experiments with more executors than physical nodes, we ran several single-core executors inside Docker containers on each worker node.  This way the executors cannot share JVM memory and behave independently.
%We take care that containers have the required resources for the Spark executors.  
We used 12 nodes as worker nodes, and 1 node as the master and driver node.  The worker nodes also hosted a Hadoop Distributed File System (HDFS) to store the experiment logs.

%\begin{table}[hb!]
%    \centering
%    \begin{tabular}{ | c | c | } 
%        \hline
%        No. of physical nodes & 13 \\ 
%        \hline
%        CPU & Intel Xeon Silver 4110 \\ 
%        \hline
%        Memory & 16 GB DDR4 2666MHz \\
%        \hline
%        Storage & Crucial MX500 \\
%        \hline
%        Network & 1 Gbit Switch \\
%        \hline
%        Distributed File System & HDFS 2.7.2 \\
%        \hline
%        Apache Spark Version & 3.0.1 modified \\
%        \hline
%        Docker version & 19.03.1 \\
%        \hline \hline
%        Executor cores & 1 \\
%        \hline
%        Executor memory & 1 GB \\
%        \hline
%        Driver cores & 4 \\
%        \hline
%        Driver memory & 8G \\
%        \hline
%    \end{tabular}
%    \caption{Hardware and Software used for the Apache Spark experiments.}
%    \label{tab:hardware_configuration}
%\end{table}

We used Apache Spark version 3.0.0 preview2, modified slightly to log more details about the processing of tasks and jobs~\cite{bora:spark-logging}.  Specifically we added a Spark listener which stores more detailed task metrics than what is available by default.  To run the experiments we extended SparkBench~\cite{10.1007/978-3-319-31409-9_3}, a tool created to benchmark Spark clusters.  We implemented classes which make it possible to run workloads in the manner of a split-merge or single-queue fork-join system.  We also added workloads to create jobs composed of tasks with service times sampled from known distributions.
%The tasks on the executors run for the specified time and sit in a loop, multiplying numbers, to create actual load on the physical node.

Our environment makes it possible to perform experiments on a Spark cluster that match as closely as possible the statistical assumptions about task service time distributions made later in developing an analytical model for tiny tasks.  This is essential when validating our analytical results through experiments and simulation.

%
%------------------------------------------------------------------------
%------------------------------------------------------------------------
%
\subsection{Simulation}
\label{sec:simulation}
We ran additional experiments using \texttt{forkulator}, an event-driven simulator for parallel systems~\cite{forkulator}.  Because it is not constrained to running in real time, using a simulator allows us to generate orders of magnitude more data points than the Spark experiments.  It also lets us run idealized experiments both with and without the scheduling and processing overhead inherent to real experiments, and allows us to experiment with the effects of different types of overhead and overhead distributions to see which most accurately matches the behavior of real systems.

%Forkulator is written in Java.  It contains classes that extend its \texttt{IntertimeProcess} interface and can be instantiated and used as arrival and/or task service time processes.  We further extended the \texttt{ExponentialIntertimeProcess} class to allow the addition of different types of overhead to the distribution.  We discovered that an overhead distribution that is the sum of a constant component and an exponentially distributed component best matches the behavior of our experimental Spark cluster.  The details of the experiments will be given in the following sections.
%
%{\color{red} say more about this...?}

%
%------------------------------------------------------------------------
%------------------------------------------------------------------------
%

\subsection{Evaluation}
\label{sec:evaluation}

We configured the Spark experiments with $l = 50$ workers, iid exponential inter-arrival times with parameter $\lambda = 0.5 \, \text{s}^{-1}$.  Initially we create jobs with no tinyfication, ${k = l = 50}$ tasks with iid exponential task service times with mean 1000~ms.  With tinyfication we take $k > l$ tasks per job with iid task service times drawn from an exponential distribution with  parameter $\mu = \frac{k}{l} \,\text{sec}^{-1}$.  This means that as the number of tasks per job, $k$, increases, the mean service time of the tasks will correspondingly decrease, keeping the expected job workload, $\text{E}\left[L(n)\right] = k/\mu = l \, \text{s}$, constant.  
For each configuration we submitted at least 30,000 jobs.  The jobs were run in batches of 1000 to ensure the accumulated measurements of each batch could be stored without interfering with the experiment.

% Can be removed if we say the output of the exponential distribution is interpreted in seconds
%The \textbf{sojourn time improvement} {\color{red} (why bold?)} of tiny tasks can be seen in
The performance benefit of using tiny tasks in both split-merge and single-queue fork-join systems can be seen in Fig.~\ref{fig:sojourntime_sim_vs_spark} which plots the $\varepsilon=0.99$ quantile of job sojourn time.
For the fork-join system in Fig.~\ref{fig:sojourntime_sim_vs_spark_fj}, a 12-fold tinyfication (going from $k=50$ to $k=600$ tasks per job), decreases the sojourn time quantile by 46.7\%.  The benefit is most dramatic at smaller tinyfication factors.  Going from $\kappa=1$ to $\kappa=2$ alone (from $k=50$ to $k=100$), reduces the sojourn time quantile by 30.4\%.

%
% tinyfication on fork-join 0.99 sojourn time quantile
% k=50  9398.2
% k=100 6543.740
% k=200 5458.600
% k=300 5130.0
% k=400 5210.20
% k=500 5073.29
% k=600 5005.60
% k=800 5071.80
% k=1000    5076.80
%

The results for the split-merge system are shown in Fig.~\ref{fig:sojourntime_sim_vs_spark_sm}.  For the arrival and service parameters we used, the split-merge system is unstable in the big tasks case ($\kappa=1$).  This is expected based on~\cite[Eq. 21]{rizk:forkjoin} and the stability analysis that we will present in Sec.~\ref{sec:smtt-stability}.
A four-fold tinyfication, splitting the jobs into $k=200$ tiny tasks, stabilizes the system.  More extreme tinyfication reduces the sojourn times further.

% Looking at the split-merge system also the \textbf{stabiltiy improvement} can be noticed. The job sojourn time is growing over time for low tinyfication factors like $\kappa = 2$ and therefor the system must be considered unstable, a factor of $\kappa = 4$ results in a stable system. This corresponds to the stability region of our simulations in Fig.~\ref{fig:splitmergestabilitytinytaskskappastability}. \todo{Maybe we need to discuss if my way of deciding if the experiment is stable is valid}
% The figures show the huge impact of tiny tasks in the sojourn time and is in line with the expectation based on Fig.~\ref{fig:sojourntimeconvergence}.

The limits of tiny tasks start to become apparent at higher tinyfication factors, $\kappa$.  In this experiment, beyond about ${k=1000}$ tasks per job, the gains start to level off.  With an increasing number of smaller and smaller tasks
%, the ratio of the overhead compared to the task service times, $O_i(n)/Q_i(n)$, gets worse, and 
the sojourn time quantiles begin to increase.  
%We see that this effect appears at lower tinyfication factors in the fork-join than the split-merge system.  This is because the performance gain from using tiny tasks in the split-merge case is so much more drastic (reducing idle times) that it outweighs the cost of overhead from extreme tinyfication.  Because the fork-join system has no end barriers, and less idle time to start with, the performance gains from tiny tasks are smaller, and cost of the scheduling overhead becomes apparent at smaller $\kappa$.

\begin{figure}
    \centering
    \subfigure[Fraction of task service time due to overhead]{
    \includegraphics[width=0.95\linewidth]{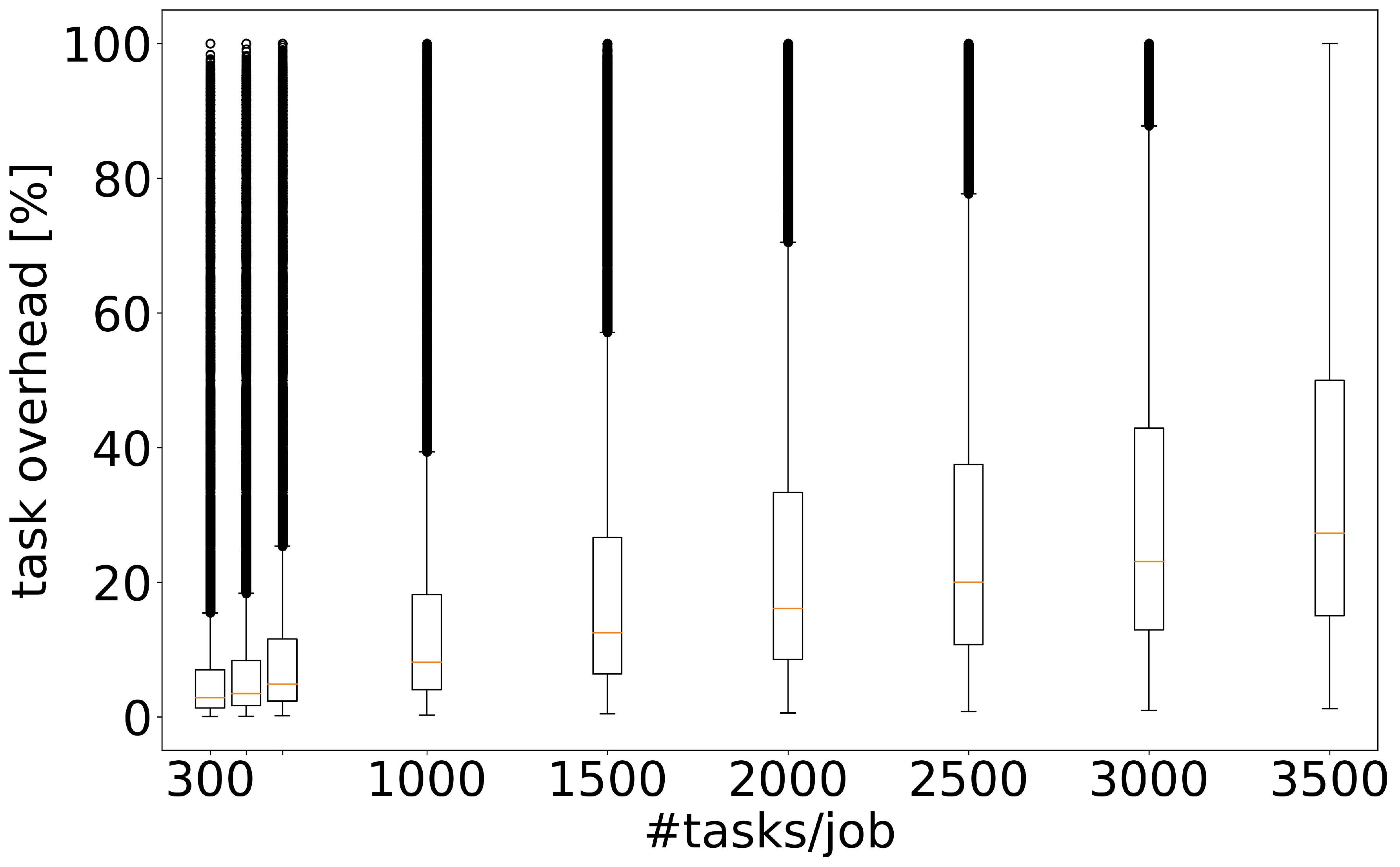}
    \label{fig:task_overhead_relation}
    }
    \subfigure[Total task overhead per job.]{
    \includegraphics[width=0.95\linewidth]{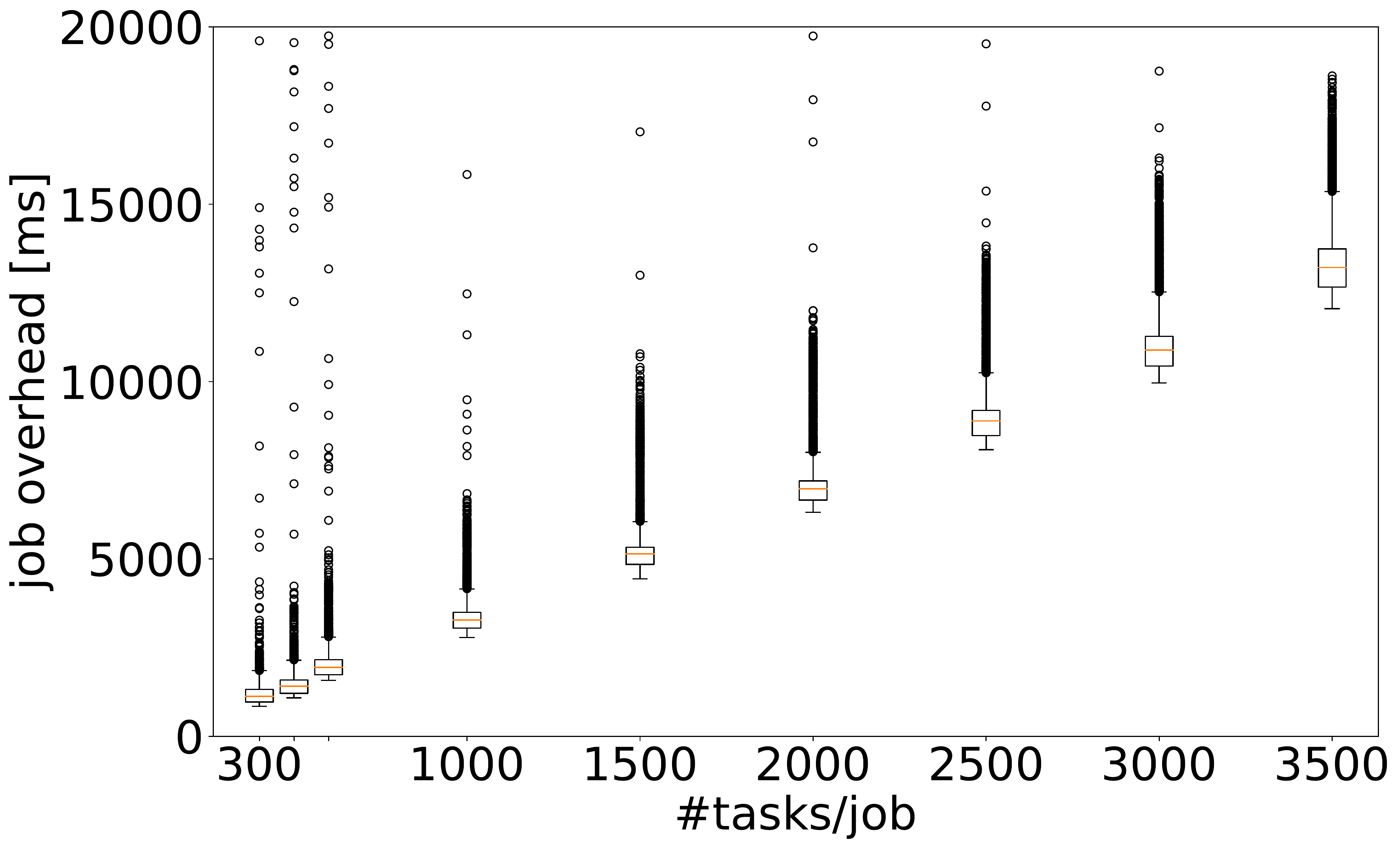}
    \label{fig:job_overhead_cropped}
    }
    \caption{Overhead on the example of the fork-join experiments with the configuration of Fig.~\ref{fig:sojourntime_sim_vs_spark}.}
    \label{fig:overhead}
\end{figure}

%Even Fig.~\ref{fig:detailed_task_overhead} is showing that fetching the serialized task is a major part of the overhead while executing the first task of a stage per executor the overhead is considered without this distinction in the following. With the increase of task per job this overhead can be neglected.

To investigate this in more detail we look at the fraction of the total task service time that is due to overhead at different tinyfication factors.  Fig.~\ref{fig:task_overhead_relation} shows a box plot of $O_i(n)/Q_i(n)$ vs. $k$ for the split-merge execution mode.  Both the median and mean of the overhead fraction grow nearly linearly with increasing $k$.  We see that there are outlier tasks which exhibit close to 0\% and others close to 100\% overhead.  This is mainly due to the random task service time distribution which can produce large or very small service times.
%If we assume a small constant overhead, $O_i(n)=c$, then tasks with an execution time close to zero will consist of nearly 100\% overhead, while for large outlier tasks the same constant overhead will account for a negligible fraction of the task service time.
Fig.~\ref{fig:job_overhead_cropped} shows a boxplot of the total overhead per job, $O(n)=\sum_{i=1}^{k}O_i(n)$, for a range of $k$ values.  The median shows a nearly linear increase with $k$.  This growth in total overhead partially explains the growth in sojourn times observed in Fig.~\ref{fig:sojourntime_sim_vs_spark}.  The form and rate of this increase will depend on understanding the distribution of the overhead, and its affect on job waiting times, which we will explore throughout the rest of this paper.
% NOTE: just knowing that the median total overhead per job increases linearly isn't enough to say what the effect on sojourn time quantiles will be.  It it more accurate to say "partially explains", because knowing this definitely does not fully predict the results we see.
%of the experimental results after a specific tinyfication factor instead of converging to the perfect partitioning.

%
%------------------------------------------------------------------------
%------------------------------------------------------------------------
%

\subsection{Overhead distribution}
\label{sec:overhead-distribution}

We observe that the task overhead has one or more random components, and we would like to characterize its distribution so it can be modeled and replicated in simulation.  We will evaluate the accuracy of our overhead model using PP plots of the resulting job sojourn time distributions from simulation against those from Spark experiments.

\begin{figure}
\centering
\includegraphics[width=0.95\linewidth]{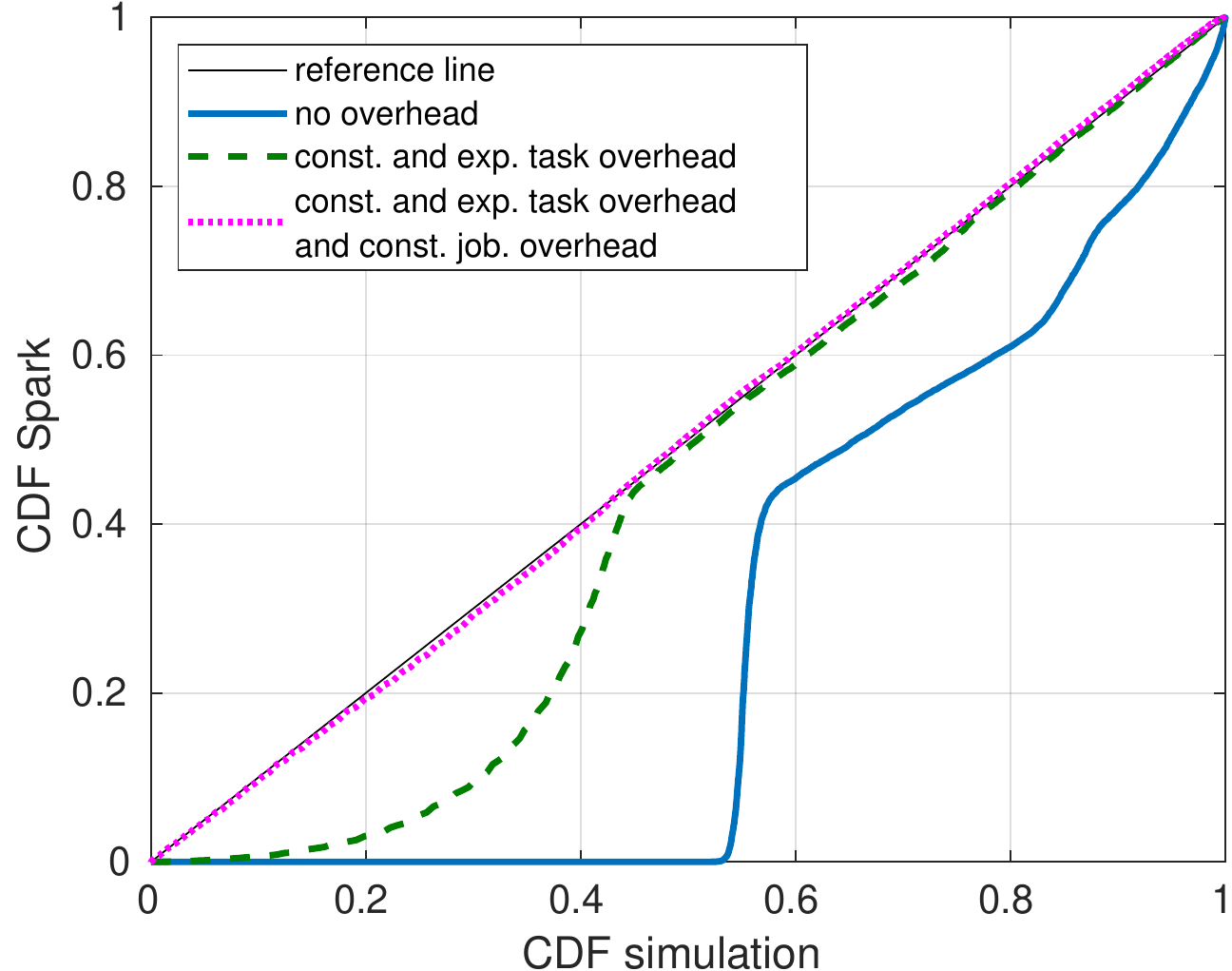}
\caption{Comparison of the single-queue fork-join sojourn time from Spark experiments and simulations. The configuration is like in Fig.~\ref{fig:sojourntime_sim_vs_spark} for the fork-join model with $k = 2500$ tasks per job. The task time overhead is added during the simulation and effects the service time. The job overhead is added after the simulation and can be interpreted as an asynchronous overhead of the spark scheduler while the next job can already run.}
\label{fig:pp_plot}
\end{figure}

Fig.~\ref{fig:pp_plot} shows the PP plot of the job sojourn times of a simulated single-queue fork-join system with exponential task execution times and $k=2500$ tasks per job, both with and without simulated overhead, against those of the equivalent Spark experiments.
The blue line is the PP plot when no overhead is included in the simulation.  It shows that the CDF of Spark sojourn times remains at or close to zero while at least half of the simulated jobs depart.  After that the Spark CDF catches up gradually.  A step-like pattern in a PP plot means that the support of one of the distributions is offset by some amount.  Based on this, and the linear growth of the job overhead in Fig.~\ref{fig:job_overhead_cropped}, we added a constant amount of overhead, $c^{ts}_{task}$, to every task in the simulation.  We also observe a scattering of extreme outliers in the overhead.  We model them by adding an additional exponential component to the task overhead with mean $0.5 \ \text{ms}$ ($\mu^{ts}_{task} = 2000 \ \text{s}^{-1}$).  This gives us a two-parameter model for task-service overhead.
\begin{equation}
    O_i(n) \sim c^{ts}_{task} + \text{Exp}(\mu^{ts}_{task})
    \label{eq:task-overhead-exponential}
\end{equation}
%These types of overhead that block subsequent tasks, and can be modeled as an addition to the task service time distribution, are what we classified as task-service overhead.

The resulting PP is plotted with a dashed green line in Fig.~\ref{fig:pp_plot}.  The two distributions fit each other much better starting at around 50\% of the samples.  This can be interpreted to mean that the minimal sojourn times in the Spark experiment are higher compared to the simulation.  We hypothesize that this is because of the processing time on the driver application, and simulated it by adding some amount, $c^{pd}_{job}$, of pre-departure overhead to every job.  The amount of pre-departure overhead needed turns out to grow linearly with the number of tasks, with rate $c^{pd}_{task}$.  %The $y$-intercept of this line is a constant amount of pre-departure overhead added per job.  
We model this overhead with a deterministic linear function added to the simulated departure time.
\begin{equation}
    D^o(n) = D(n) + c^{pd}_{job} + k \cdot c^{pd}_{task}.
    \label{eq:predeparture-overhead}
\end{equation}

In the fork-join case, this did not require modifying the simulation, since the pre-departure overhead is simply added to the simulated sojourn times.  It does not affect the processing of subsequent jobs or tasks.  In the split-merge case, delaying the departure of the job does block the tasks of subsequent jobs, and therefore did require modifications to the corresponding scheduler class in the simulator.
%\footnote{The assumptions that this per-job overhead is constant, and the assumption of inter-job independence are not completely accurate.  In reality there are shared queues and housekeeping mutexes in the scheduler which could add a tiny bit of variation and interdependence, especially under high job load.}
%Our simulations show that the constant overhead of the sojourn time increases with the number of tasks per job.  

\begin{center}
\bgroup
\def\arraystretch{1.3}
\begin{tabular}{| c | r c l |} 
 \hline
 \multirow{2}{*}{ task-service } & $c^{ts}_{task}$ & = & $2.6 \ \text{ms}$ \\
 \cline{2-4}
      & $\mu^{ts}_{task}$ & = & $2000 \ \text{s}^{-1}$ \\
 \hline
 \multirow{2}{*}{ pre-departure } & $c^{pd}_{job}$ & = & $20 \ \text{ms}$ \\
 \cline{2-4}
      & $c^{pd}_{task}$ & = & $7.4 \times 10^{-3} \ \text{ms}$ \\
 \hline
\end{tabular}
\egroup
\end{center}
%Constant overhead per job (non destructive): 0.024s
%Constant overhead per task (non destructive): 7.4e-06s
%Constant overhead per task (destructive): 0.0026s
%Exponential overhead per task (destructive): Parameter 2000, so the expected value is 0.0005s 

The parameter values determined in these experiments are listed in the table above.  The magenta dotted line in Fig.~\ref{fig:pp_plot} shows the PP plot of the simulation with both constant and exponential task overhead and linear pre-departure overhead against the Spark result.  
In comparison to the black reference line, simulation with these three overhead components, matches the distribution of the real Spark experiment acceptably well.  Simulating an overhead distribution for the first task of each job on each executor, as might be expected based on Fig.~\ref{fig:detailed_task_overhead}, turns out not to be necessary to make these distributions match.

The dashed lines in Fig.~\ref{fig:sojourntime_sim_vs_spark} show the corresponding sojourn time quantiles from simulations with this overhead model.  For both the fork-join and split-merge models, the simulations with overhead match the Spark experiments very reasonably.  
%Importantly, they both match at the point at which overhead begins to overtake the performance gains due to tinyfication.
% The result visualizes that the sojourn time distribution of the simulation fits the one from the spark experiment. Therefor it seems valid to model the overheads of spark with a constant and exponential task and a constant job overhead. A distinction between the first task per executor and the other tasks indicated by the different overheads shown in Fig.~\ref{fig:detailed_task_overhead} is not necessary.
% even the overhead on a real system are never constant and also the execution model of spark generates different overheads like shown in Fig.~\ref{fig:detailed_task_overhead}.
% Also the real overhead is not constant and there can be outliers with larger overhead for example through an over utilized driver or network, it seems valid to consider constant task and job overhead to model the overheads in spark.

\begin{figure}
    \centering
    \includegraphics[width=0.95\linewidth]{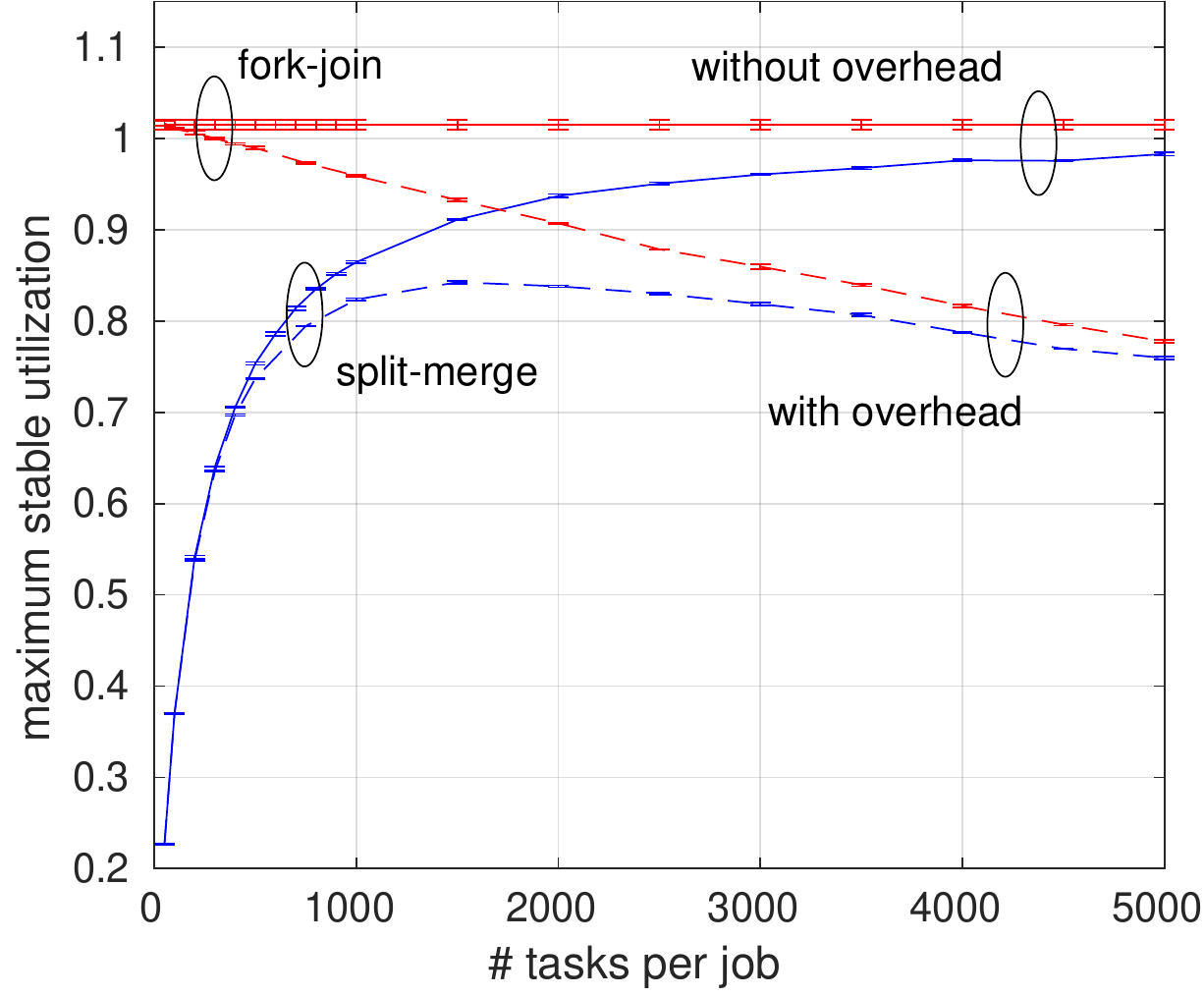}
    \caption{The stability regions of split-merge and fork-join simulated with and without task and job overhead, with $l=50$ parallel workers.}
    \label{fig:stability_with_overhead}
\end{figure}

Fig.~\ref{fig:stability_with_overhead} shows the simulated stability regions as a function of tasks-per-job, $k$, for both split-merge and fork-join systems.  The solid lines are for simulation without overhead and the dashed lines with the simulated overhead model discussed above.
The maximum stable utilization in the split-merge system is dramatically increased by using tiny tasks, but begins dropping again around 2000 tasks per job ($\kappa\approx36$) due to overhead.
The fork-join system is stable up to a utilization of 1.0 in general, so no stability improvement is possible.  In fact, the overhead reduces the stability region gradually as the tinyfication factor increases.
%The lower sojourn quantiles in Fig.~\ref{fig:sojourntime_sim_vs_spark} are explainable by the reduction of stragglers.
% Fig.~\ref{fig:sojourntime_sim_vs_spark} compares the sojourn time of the simulation with the real spark experiments with $l = 50$ servers and for different tasks $k$ per job. We show both the single-queue fork-join model in red and the split-merge model in blue. The error bars represents the 95\% confidence interval. The course of the sojourn time of our spark experiments follow the simulated one in both models until the size of the overhead exceeds the advantage of the tinyfication.

%
%------------------------------------------------------------------------
%------------------------------------------------------------------------
%------------------------------------------------------------------------
%

\section{Network calculus for parallel computing systems}
\label{sec:anaytical-models}
We start by presenting the analytical tools and notation needed to derive and understand our analytical models of tiny tasks.  We build on the approach of~\cite{rizk:forkjoin, fidler:multiserver}, making use of a max-plus version of the stochastic network calculus.
%The analytical models have to assume iid exponential task service times for reasons that we will highlight as they come up.  That is, we model the system without the variety of additional overhead distributions discussed in section~\ref{sec:overhead-distribution}.  These models, therefore, help us understand the fundamental performance gains possible using tiny tasks, but not the performance trade-off due to scheduling overhead.  At least not directly.
%
%
%------------------------------------------------------------------------
%
\subsection{Notation and background}

We use the same naming and notation for the random processes that were discussed in Sec.~\ref{sec:systems-models-stability}, and introduce some additional items.  
As before $A(n)$ and $D(n)$ for $n \ge 1$ denote the {\bf arrival time} and {\bf departure time} of job $n$, and $Q_i(n)$ denotes the {\bf task service time} of task $i \in [1,k]$ of job $n$.  By convention we take $A(0)=0$.  $A(m,n) = A(n)- A(m)$ is the {\bf inter-arrival time} between jobs $n \ge m$.  $L(n)=\sum_{i=1}^{k}Q_i(n)$ denotes the (total) {\bf workload} of the job.  The {\bf job service time} $\Delta(n)$ is the total time a job spends in service.  Note that for the parallel models, $L(n)$ and $\Delta(n)$ are generally not equal.

Servers are modeled using a definition of {\bf max-plus server} with {\bf service process} $S(m,n)$, adapted from~\cite[Def. 6.3.1]{chang:performanceguarantees}.  
\begin{definition}[Max-plus server]
\label{def:maxplusserviceprocess}
A system with arrivals $A(n)$ and departures $D(n)$ is an $S(m,n)$ server under the max-plus algebra if it holds for all $n \ge 1$ that
\begin{equation*}
D(n) \le \max_{m \in [1,n]} \{ A(m) + S(m,n) \} .
\end{equation*}
\end{definition}

Applying this definition to the {\bf sojourn time}, $T(n) = D(n)-A(n)$ for $n \ge 1$, we obtain:
\begin{equation}
T(n) \le \max_{m \in [1,n]} \{S(m,n) - A(m,n)\} .
\label{eq:sojourntime}
\end{equation}
In the case of first-come first-served service, an expression for the {\bf waiting time} $W(n)=[D(n-1)-A(n)]_+$, where $[X]_+ = \max\{0,X\}$, can be derived in the same way.

In the case of single-server systems, the service process corresponds to the cumulative service time of jobs $m$ to $n$ and we have the relationship $S(m,n)=\sum_{\nu=m}^{n} \Delta(\nu)$ where $\Delta(\nu)$ is the service time of job $\nu$.  When we move to the multi-server setting, the definition of $S(m,n)$ becomes more subtle.  For example $S(m,n)$ may not generally be defined in increments of $\Delta(n)$.

Just as in~\cite{rizk:forkjoin, fidler:multiserver} we will make use of {\bf moment generating functions} (MGFs) of the arrival and service processes.  The MGF of a random variable $X$ is defined as $\mathsf{M}_X(\theta) = \mathsf{E}[e^{\theta X}]$ where $\theta$ is a free parameter.  The MGF has the properties that $\mathsf{M}_{X+Y}(\theta) = \mathsf{M}_{X}(\theta) \mathsf{M}_{Y}(\theta)$ for  $X$ and $Y$ independent, and that $\mathsf{M}_{cX}(\theta) = \mathsf{M}_{X}(c\theta)$ for any constant, $c$.

A common class of MGF models are {\bf$(\sigma,\rho)$-envelopes} defined in~\cite[Def. 7.2.1]{chang:performanceguarantees}.  These are adapted to max-plus servers in~\cite[Def. 2]{fidler:multiserver}.
\begin{definition}[$(\sigma,\rho)$-Arrival and Service Envelopes]
\label{def:sigmarho}
An arrival process, $A(m,n)$, is $(\sigma_A,\rho_A)$-lower constrained if for all $n \ge m \ge 1$ and $\theta > 0$ it holds that
\begin{equation*}
\mathsf{E}\Bigl[e^{-\theta A(m,n)}\Bigr] \le e^{-\theta (\rho_A(-\theta) (n-m) - \sigma_A(-\theta))} .
\end{equation*}

A service process, $S(m,n)$, is $(\sigma_S,\rho_S)$-upper constrained if for all $n \ge m \ge 1$ and $\theta > 0$ it holds that
\begin{equation*}
\mathsf{E}\Bigl[e^{\theta S(m,n)}\Bigr] \le e^{\theta(\rho_S(\theta) (n-m+1) + \sigma_S(\theta))} .
\end{equation*}
\end{definition}
Max-plus servers with $(\sigma,\rho)$-envelopes are models of G$\mid$G$\mid$1 queues, and a variety of stochastic processes satisfy the definition including Markov and periodic processes~\cite{chang:performanceguarantees, kelly:effectivebandwidths, fidler:netcalcsurvey}. In this work we restrict ourselves to GI$\mid$GI$\mid$1 queues.
%with independent and identically distributed (iid) increments{color{red} (redundant?)}.  For the arrival process $A(m,n) = \sum_{\nu=m}^{n-1} A(\nu,\nu+1)$ this means we assume iid inter-arrival times $A(\nu,\nu+1)$, and for the service process $S(m,n)$ the individual service times of each job $\Delta(n)$ are iid.  
In the iid case we have $\sigma_A(-\theta)=\sigma_S(\theta)=0$.

As an example, consider the classical M$\mid$M$\mid$1 queue.  The arrival process has iid inter-arrival times $A(n,n+1)\sim\text{Exp}(\lambda)$, and MGF $\mathsf{E}[e^{-\theta A(n,n+1)}] = \lambda/(\lambda + \theta)$ for $n \ge 1$ and $\theta > 0$. It follows that
\begin{equation}
\rho_A(-\theta) = - \frac{1}{\theta} \ln \left(\frac{\lambda}{\lambda+\theta}\right) ,
\label{eq:rhoexpoarrivals}
\end{equation}
for $\theta > 0$. Similarly, for iid service times $\Delta(n) \sim \text{Exp}(\mu)$ we have $\mathsf{E}[e^{\theta \Delta(n)}] = \mu/(\mu - \theta)$ for $n \ge 1$ and $\theta \in (0,\mu)$ so that
\begin{equation}
\rho_S(\theta) = \frac{1}{\theta} \ln \left(\frac{\mu}{\mu-\theta}\right) ,
\label{eq:rhoexposervice}
\end{equation}
for $\theta \in (0,\mu)$. In this example parameter $\rho_A(-\theta)$ decreases with $\theta > 0$ from the mean inter-arrival time to the minimal inter-arrival time (possibly zero) and $\rho_S(\theta)$ increases with $\theta > 0$ from the mean service time to the maximal service time (possibly infinity).

Performance bounds are obtained using a basic theorem of the stochastic network calculus, e.g.,~\cite[Th. 1]{fidler:multiserver}.

\begin{theorem}[Statistical sojourn time bound]
\label{th:sojourntimesingleserver}
Given an $S(m,n)$ server with iid inter-arrival times with envelope rate $\rho_A(-\theta)$ and iid service times with envelope rate $\rho_S(\theta)$. For any $\theta > 0$ that satisfies $\rho_S(\theta) \le \rho_A(-\theta)$, the waiting time for all $n \ge 1$ is bounded by
\begin{equation*}
\mathsf{P}[W(n) > \tau] \le e^{-\theta \tau},
\end{equation*}
and the sojourn time by
\begin{equation*}
\mathsf{P}[T(n) > \tau] \le e^{\theta \rho_S(\theta)} e^{-\theta \tau}.
\end{equation*}
\end{theorem}
%
%Similar results are available for the backlog and for the case of non-iid arrival and service processes. The free parameter $\theta$ can be optimized. For the example of the M$\mid$M$\mid$1 queue the maximal speed of the tail decay $\theta$ follows from $\rho_S(\theta) \le \rho_A(\theta)$ as $\theta = \mu-\lambda$ so that $\mathsf{P}[T(n) > \tau] \le \frac{\mu}{\lambda} e ^{-(\mu-\lambda)\tau}$.

%
%------------------------------------------------------------------------
%------------------------------------------------------------------------
%
\subsection{State of the art in parallel systems}
Here we will summarize prior results for split-merge, fork-join, single-queue fork-join, and ideal partitioning parallel systems for the ``big-tasks'' case, where the number of tasks per job, $k$, equals the number of servers, $l$~\cite{rizk:forkjoin}.
%In the parallel setting we need to define additional terms and notation to distinguish between the service processes for jobs and tasks.
%Also we define notation for the more tangible properties of the job service.

%

%
%------------------------------------------------------------------------
%------------------------------------------------------------------------
%
\subsubsection{Split-merge} In the big-tasks split-merge model all tasks in a job start simultaneously.  Therefore the system can be modeled like a single-server system where each job's service time is determined by that of its maximal task $\Delta(n) = \max_{i \in [1,l]} \{Q_i(n)\}$.  Hence, for $n \ge m \ge 1$ the model can be expressed as a max-plus server with service process~\cite{rizk:forkjoin,fidler:multiserver} 
\begin{equation}
S(m,n) = \sum_{\nu=m}^n \max_{i \in [1,l]} \{Q_i(\nu)\} .
\label{eq:splitmergeserver}
\end{equation}

For iid $Q_i(n)\sim \text{Exp}(\mu)$ it also follows that the service process of the split-merge model~\eqref{eq:splitmergeserver} has service envelope
\begin{equation}
\rho_S(\theta) = \frac{1}{\theta} \sum_{i=1}^l \ln \left( \frac{i\mu}{i\mu-\theta} \right),
\label{eq:rhosplitmerge}
\end{equation}
for $\theta \in (0,\mu)$~\cite{rizk:forkjoin}. The sojourn time bound depicted in Fig.~\ref{fig:sojourntimecomparison} is obtained by substitution of~\eqref{eq:rhosplitmerge} into Th.~\ref{th:sojourntimesingleserver} and optimizing subject to $0 <\theta<\mu$.
%

%
%------------------------------------------------------------------------
%------------------------------------------------------------------------
%
\subsubsection{Fork-join} The service process of the fork-join model is
\begin{equation}
S(m,n) = \max_{i \in [1,l]} \left\{ \sum_{\nu=m}^n Q_i(\nu) \right\} ,
\label{eq:forkjoinserver}
\end{equation}
for $n \ge m \ge 1$~\cite{rizk:forkjoin}.  This says that $S(m,n)$ is determined by the maximal sequence of tasks that are assigned to a server.  Clearly, for a given set of task service times $Q_i(n)$, the service process $S(m,n)$ of the fork-join model~\eqref{eq:forkjoinserver} will be less than or equal to that of the split-merge model~\eqref{eq:splitmergeserver}.  The sojourn time can be obtained from~\eqref{eq:sojourntime} by substitution of~\eqref{eq:forkjoinserver}, substitution of $Q_i(m,n) = \sum_{\nu=m}^n Q_i(\nu)$, and reordering of the maxima to give
\begin{equation*}
T(n) \le \max_{i \in [1,l]} \left\{ \max_{m \in [1,n]} \{ Q_i(m,n) - A(m,n)\} \right\}.
\end{equation*}
Then $T_i(n) = \max_{m \in [1,n]} \{ Q_i(m,n) - A(m,n)\}$ are the individual task sojourn times at server $i \in [1,l]$.  For each server $i \in [1,l]$ Th.~\ref{th:sojourntimesingleserver} can be used to derive $\mathsf{P}[T_i(n) > \tau]$ and applying the union bound,~\cite{rizk:forkjoin, fidler:multiserver} gives us $\mathsf{P}[T(n) > \tau] \le \sum_{i=1}^l \mathsf{P}[T_i(n) > \tau]$. The same steps can be used to derive a waiting time bound.  For the homogeneous case it follows from Th.~\ref{th:sojourntimesingleserver} that
\begin{equation*}
\mathsf{P}[T(n) > \tau] \le l e^{\theta \rho_Q(\theta)} e^{-\theta \tau},
\end{equation*}
for any $\theta > 0$ satisfying $\rho_Q(\theta) \le \rho_A(-\theta)$.  For the case of iid exponential inter-arrival and task service times, we can use $\rho_A(-\theta)$ from~\eqref{eq:rhoexpoarrivals} and substitute $\rho_S(\theta)$ from~\eqref{eq:rhoexposervice} for $\rho_Q(\theta)$ to obtain the fork-join sojourn time bound plotted in Fig.~\ref{fig:sojourntimecomparison}.

Since in this case $\rho_Q(\theta)$ and $\rho_A(-\theta)$ converge towards the mean task service time and the mean inter-arrival time, respectively, as $\theta \rightarrow 0$, the condition $\rho_Q(\theta) \le \rho_A(-\theta)$ implies that the fork-join model is stable up to a utilization of one.
%, as seen in Fig.~\ref{fig:splitmergestability} {\color{red} (get rid of this figure - redundant)}.  

%
%------------------------------------------------------------------------
%
\subsubsection{Single-queue fork-join} The service process of the single-queue fork-join model is more involved. The corresponding results in Fig.~\ref{fig:sojourntimecomparison} are obtained from~\cite[Th. 4]{fidler:multiserver}. The single-queue fork-join model is also a special case (for $k=l$) of Th.~\ref{th:tinytaskforkjoin} in this paper.

%
%------------------------------------------------------------------------
%
\subsubsection{Ideal partition} 
%The sum of $l$ iid exponential random variables has an Erlang-$l$ distribution.  Therefore 
If jobs are composed of $k$ iid exponential tasks with parameter $\mu$, then the jobs' total workload has distribution $L(n)\sim \text{Erlang}(k,\mu)$.  If jobs with this workload distribution were instead divided into $l$ equally-sized tasks, then the tasks would have an $\text{Erlang}(k,l\mu)$ distribution, so that
%
%Finally, we note that $l$ iid exponential tasks with parameter $\mu$, as used for the comparison in Fig.~\ref{fig:sojourntimecomparison}, correspond to an Erlang-$l$ job size distribution with parameter $\mu$ (The sum of $l$ iid exponential random variables is Erlang-$l$ distributed). For the results labeled 'ideal partition' in Fig.~\ref{fig:sojourntimecomparison}, we use the same job size distribution, however, partition each job into $l$ equisized tasks. It follows that the tasks are Erlang-$l$ with parameter $l\mu$ so that
%
\begin{equation}
\rho_Q(\theta) = \frac{k}{\theta} \ln \left(\frac{l\mu}{l\mu-\theta}\right)
\label{eq:rhoidealdivision}
\end{equation}
for $\theta \in (0,l\mu)$. Since the tasks of each job are equisized, all tasks of each job start and finish in unison.  Hence, the system functions identically to a single server.  The sojourn time bound depicted in Fig.~\ref{fig:sojourntimecomparison} follows by substitution of~\eqref{eq:rhoidealdivision} into Th.~\ref{th:sojourntimesingleserver}.
%
%------------------------------------------------------------------------
%
\section{Split-merge systems with tiny tasks}
\label{sec:stabilizing-splitmerge}

In this section, we extend the split-merge model to cases with finer task granularity, to understand how using tiny tasks extends its stability region and improves its sojourn time.  As before we assume $l$ workers and $k \ge l$ tasks per job.

%We model the split-merge system with tiny tasks as a max-plus server and perform an analysis for $k$ iid exponential tiny tasks with parameter $\mu$.  

%This choice corresponds to Erlang-$\kappa$ big tasks and Erlang-$k$ jobs, respectively. We show how tiny tasks remedy the stability issue of the split-merge model and how they improve the sojourn time.
%
\begin{lemma}[Tiny tasks split-merge model]
\label{lem:tinytasksplitmerge}
The split-merge model with $l$ workers and $k\ge l$ tasks per job is a max-plus server.  Given iid exponential task service times with parameter $\mu$, its service process has envelope rate $\rho_S(\theta) = \rho_X(\theta) + (k-l) \rho_Z(\theta)$, where
\begin{equation*}
\rho_X(\theta) = \frac{1}{\theta} \sum_{i=1}^l \ln \left( \frac{i \mu}{i \mu - \theta} \right),
\end{equation*}
for $\theta \in (0,\mu)$, and
\begin{equation*}
\rho_Z(\theta) = \frac{1}{\theta} \ln \left( \frac{l \mu}{l \mu - \theta} \right),
\end{equation*}
for $\theta \in (0,l\mu)$. The expected job service time is
\begin{equation*}
\mathsf{E}[\Delta(n)] = \frac{1}{\mu} \left(\frac{k}{l} +  \sum_{i=2}^l \frac{1}{i} \right).
\end{equation*}
\end{lemma}
We note that for the special case $k=l$, Lem.\ref{lem:tinytasksplitmerge} recovers the envelope rate~\eqref{eq:rhosplitmerge} and stability condition of the conventional split-merge model.  Sojourn time and waiting time bounds follow by substitution of Lem.~\ref{lem:tinytasksplitmerge} into Th.~\ref{th:sojourntimesingleserver}.

\begin{proof}
First, we show that the tiny tasks split-merge model is a max-plus server. Let $V_i(n)$ be the time task $i \in [1,k]$ of job $n \ge 1$ starts service.  Since the first $l$ tasks of a job start at the same time, we have for $i \in [1,l]$ that
\begin{equation}
V_i(n) = \max \{ A(n), D(n-1) \}.
\label{eq:tinytaskssplitmergestarttime1}
\end{equation}
For $i \in [l+1,k]$ we have
\begin{equation}
V_i(n) = V_{i-1}(n) + Z_{i-1}(n) ,
\label{eq:tinytaskssplitmergestarttime2}
\end{equation}
where $Z_{i-1}(n)$ is the time from the start of task $i-1$ of job $n$ until the next server becomes available.

We can express the departure time $D(n)$ of job $n$ relative to the start time of its last task,
\begin{equation}
D(n) = V_k(n) + X(n),
\label{eq:tinytaskssplitmergedeparturetime}
\end{equation}
where
\begin{equation}
X(n) = \max_{i \in [1,l]} \{ Y_i(n) \}
\label{eq:defmaxresidualservice}
\end{equation}
and $Y_i(n)$ for $i \in [1,l]$ are the residual service times of the tasks, including task $k$, that are in service when task $k$ starts service at $V_k(n)$.
%I.e. the last $l$ tasks of job $n$ that still have to finish service. 
By repeated substitution of~\eqref{eq:tinytaskssplitmergestarttime2} into~\eqref{eq:tinytaskssplitmergedeparturetime}, it follows that
\begin{equation*}
D(n) = V_l(n) + \left[\sum_{i=l}^{k-1} Z_i(n)\right] + X(n) .
\end{equation*}
With~\eqref{eq:tinytaskssplitmergestarttime1} this becomes
\begin{equation}
D(n) = \max \{ A(n), D(n-1) \} + \Delta(n),
\label{eq:tinytaskssplitmergerecursion}
\end{equation}
where we write the service time of job $n$ as
\begin{equation}
\Delta(n) = \left[\sum_{i=l}^{k-1} Z_i(n)\right] + X(n) .
\label{eq:tinytaskssplitmergeservicetime}
\end{equation}
By recursive insertion of~\eqref{eq:tinytaskssplitmergerecursion} we obtain 
\begin{equation*}
D(n) = \max_{m \in [1,n]} \left\{ A(m) + \sum_{\nu=m}^{n} \Delta(\nu) \right\}, 
\end{equation*}
i.e., the tiny tasks split-merge model is a max-plus server with service process $S(m,n) = \sum_{\nu=m}^{n} \Delta(\nu)$.

Next, we consider the distribution of $X(n)$ and $Z_i(n)$. Due to the memorylessness of the iid exponential task service times, the residual service times $Y_i(n)$ are also iid exponential with the same parameter $\mu$. Regarding $Z_{i}(n)$, note that when any task $i \in [l,k]$ of job $n$ starts service all servers are busy, so that the time until the next server becomes idle is the minimum of the residual service times of the $l$ tasks that are in service.  Thus $Z_i(n)$ for $i \in [l,k-1]$ is the minimum of $l$ iid exponential random variables with parameter $\mu$, and therefore the $Z_i(n)$ are iid exponential with parameter $l\mu$.

To derive the MGF of~\eqref{eq:tinytaskssplitmergeservicetime}, we apply the identity (used by the authors of~\cite{rizk:forkjoin} to compute the stability region of the split-merge model) $\max_{i \in [1,l]} \{ Y_i(n) \} =_d \sum_{i=1}^l Y_i(n)/i$ to~\eqref{eq:defmaxresidualservice}, obtaining
\begin{equation}
\mathsf{M} [X(n)](\theta) = \prod_{i=1}^l \mathsf{M} \left[ \frac{Y_i(n)}{i} \right] (\theta) = \prod_{i=1}^l \frac{i\mu}{i\mu-\theta} ,
\label{eq:tinytaskssplitmergemgfy}
\end{equation}
for $\theta \in (0,\mu)$. Also, we have
\begin{equation}
\mathsf{M} \left[ \sum_{i=l}^{k-1} Z_{i}(n) \right] (\theta) = \mathsf{M}[Z_i(n)]^{k-l} = \left(\frac{l\mu}{l\mu - \theta}\right)^{k-l},
\label{eq:tinytaskssplitmergemgfz}
\end{equation}
for $\theta \in (0,l\mu)$. The MGF of~\eqref{eq:tinytaskssplitmergeservicetime} follows as the product of~\eqref{eq:tinytaskssplitmergemgfy} and~\eqref{eq:tinytaskssplitmergemgfz}. Taking the logarithm and dividing by $\theta$ gives $\rho_S(\theta)$.

Finally, the expected value $\mathsf{E}[\Delta(n)]$ can then be derived by substituting~\eqref{eq:defmaxresidualservice} into~\eqref{eq:tinytaskssplitmergeservicetime} and using the identity~\eqref{eq:maxofexpos}. With $\mathsf{E}[Z_i(n)] = 1/(l\mu)$, and $\mathsf{E}[Y_i(n)/i] = 1/(i\mu)$ this gives us
\begin{equation*}
\mathsf{E}[\Delta(n)] = \frac{k-l}{l\mu} + \frac{1}{\mu} \sum_{i=1}^l \frac{1}{i}.
\end{equation*}
Some reordering of the terms completes the proof.
\end{proof}

Note that the step deducing that the $Y_i(n)$ are iid exponential, and thereby equation~\eqref{eq:tinytaskssplitmergemgfy}, required the assumption of iid exponential task service times.  This is one reason that we cannot directly use a task service time that includes arbitrary non-exponential overhead in the tiny tasks model.

\begin{figure*}
\subfigure[stability region]{
\includegraphics[width=0.45\linewidth]{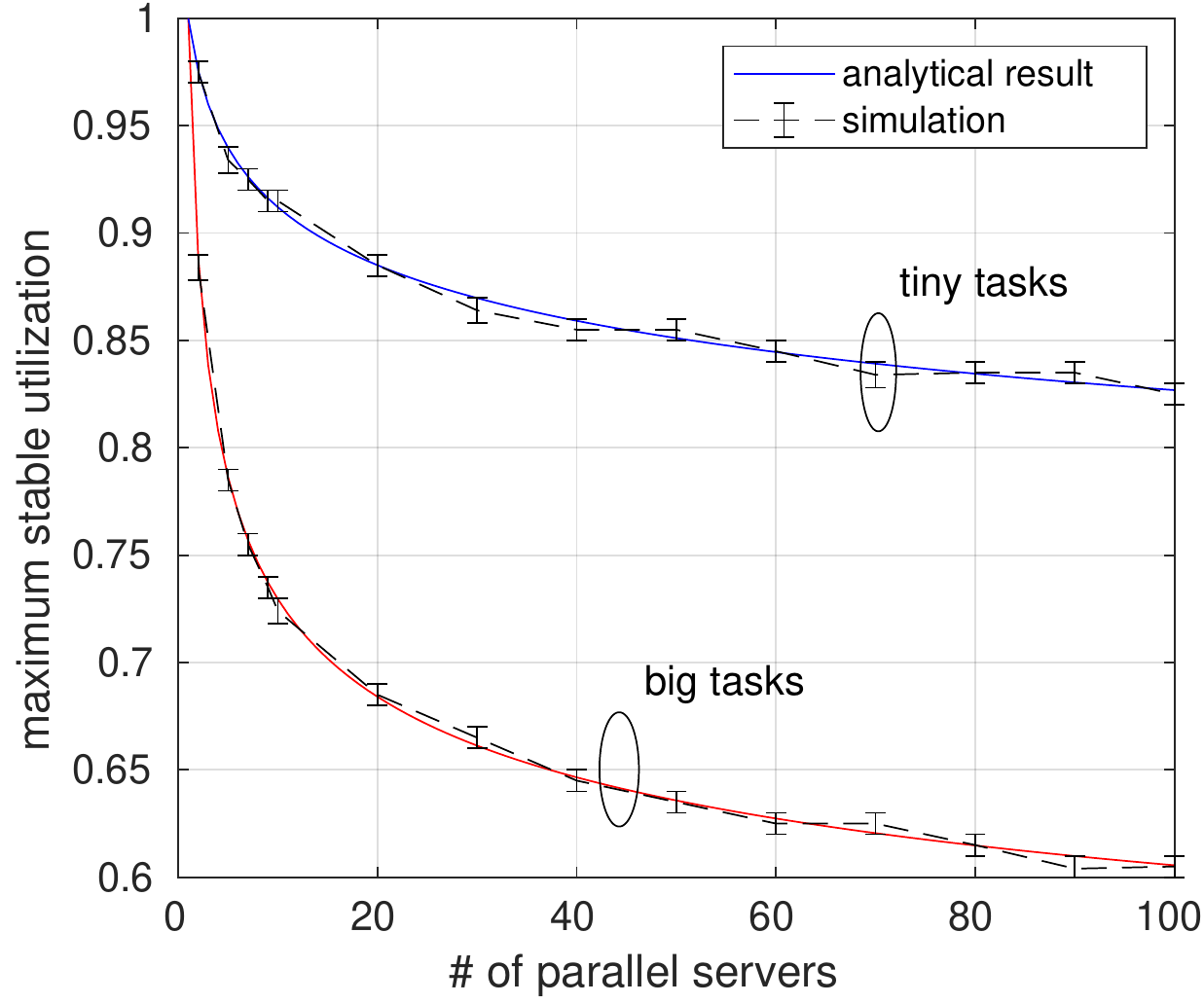}
\label{fig:splitmergestabilitytinytasks}
}
\subfigure[sojourn time $10^{-6}$ quantiles]{
\includegraphics[width=0.45\linewidth]{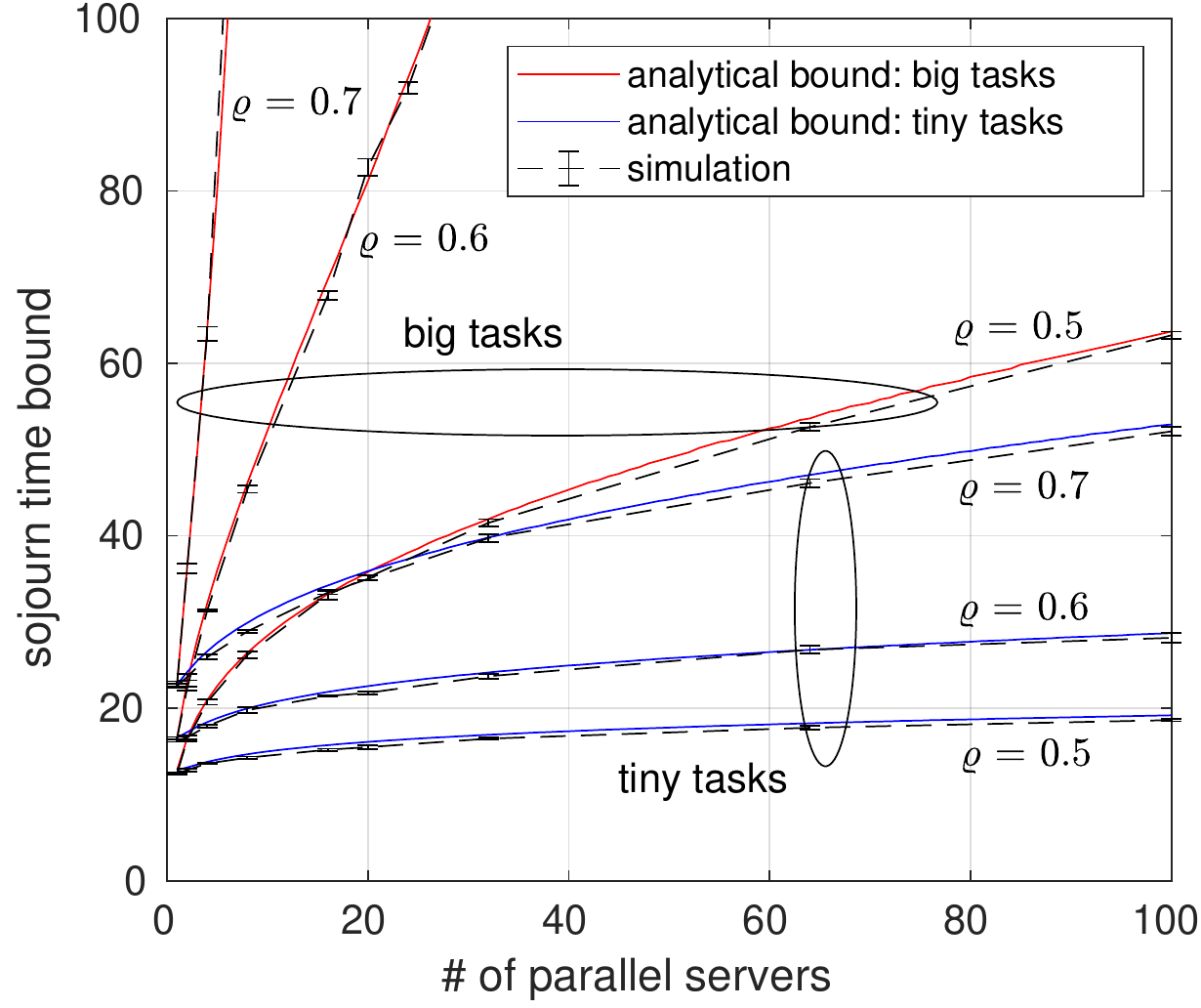}
\label{fig:sojourntimecomparisonsplitmergetinytasks}
}
\caption{Direct refinement of big tasks into tiny tasks for the split-merge model with $\text{Exp}(\lambda)$ arrivals.  Big-tasks jobs have $k=l$ $\text{Erlang}(\kappa,\mu)$ tasks.  Tiny-tasks jobs have $k=\kappa l$ $\text{Exp}(\mu)$ tasks, and are therefore a direct refinement of the corresponding big-tasks jobs.  In all plots $\mu = \kappa = 20$ so the utilization is determined by the arrival rate $\varrho = \kappa \lambda/\mu = \lambda$.
\label{fig:splitmerge-bigtasks-tinytasks}}
\end{figure*}

%
%------------------------------------------------------------------------
%------------------------------------------------------------------------
%
\subsection{Direct refinement into tiny tasks}

For most of the comparisons in this paper we fix the number of servers, $l$, and increase the number of tasks per job, $k$, making the tasks smaller to compensate; for example in Figs.~\ref{fig:sojourntime_sim_vs_spark}, \ref{fig:stability_with_overhead}, and~\ref{fig:sojourntimeconvergence}.  Another view is to fix the factor of tinyfication $\kappa=k/l$.  Then both $k$ and $l$ must increase proportionately.  Because of the relative simplicity of the $k=l$ case of the split-merge model, in this way we can make an especially direct comparison of the effects of tiny tasks, wherein the distribution of the jobs' workloads, $L(n)$, does not change.  The following will not be possible with fork-join systems, and will only be used in this section.
%, specifically in Fig.~\ref{fig:splitmerge-bigtasks-tinytasks}.  
%Generally, for comparisons where we vary the number of tasks per job, as $k$ changes, the distribution of the jobs' workloads, $L(n)$, also changes.  Even if we manipulate the service rate to hold the mean workload constant, $L(n)$ has an {Erlang-$k$} distribution, and its variance decreases as $k$ increases.

Fix some integer tinyfication factor, $\kappa$.  In the tiny tasks model we assume $k=\kappa l$ tasks per job, with $\text{Exp}(\mu)$ service times.  In the equivalent big tasks split-merge model , we have $k=l$ tasks per job with $Q_i(n)\sim\text{Erlang}(\kappa,\mu)$ service times.  In this way, the tiny tasks model is a direct refinement of the equivalent big tasks model, and importantly, the distribution of the total workload of each job stays the same.  
The key property here is that the uniformly random partitioning of $\text{Erlang}(\kappa,\mu)$ samples into $\kappa$ sub-intervals, produces iid $\text{Exp}(\mu)$ samples~\cite{steutel:random-division}, whereas random partitioning of most random variables results in non-independent sub-intervals.

\subsection{Stability of split-merge with tiny tasks}
\label{sec:smtt-stability}

To deduce the stability region of the big tasks split-merge model with iid exponential task service times, the authors of~\cite{rizk:forkjoin} use the identity
\begin{equation}
\max_{i \in [1,l]} \{Q_i(n)\} =_d \sum_{i=1}^l \frac{Q_i(n)}{i},
\label{eq:maxofexpos}
\end{equation}
where $=_d$ denotes equality in distribution.  It follows that for iid task service times $Q_i(n)\sim \text{Exp}(\mu)$, the mean job service time is $\mathsf{E}[\Delta(n)] = \sum_{i=1}^l 1/(i\mu)$.  For iid inter-arrival times $A(n,n+1)\sim \text{Exp}(\lambda)$, i.e., with mean inter-arrival time $1/\lambda$, the split-merge system is stable if and only if~\cite[Eq. 21]{rizk:forkjoin}
\begin{equation*}
\frac{1}{\lambda} > \frac{1}{\mu} \sum_{i=1}^l \frac{1}{i}.
\end{equation*}
Recall that $\varrho = \lambda/\mu$ is the utilization.  The term $\sum_{i=1}^l \frac{1}{i}$ is the $l$th harmonic number.  These have the logarithmic asymptotic limit $\gamma + \ln l$, where $\gamma \approx 0.577$ is the Euler constant. Hence, the maximum stable utilization decays proportionally to $1/\ln l$ as seen in Fig.~\ref{fig:splitmergestabilitytinytasks}.

The tiny tasks split-merge model is stable as long as the expected inter-arrival time $\mathsf{E}[A(n,n+1)]$ is larger than the expected job service time $\mathsf{E}[\Delta(n)]$.  For iid inter-arrival times $A(n,n+1)\sim\text{Exp}(\lambda)$, the condition $\lambda \mathsf{E}[\Delta(n)] < 1$ implies stability. Since the expected total workload of a job is $\mathsf{E}[L(n)]=\sum_{i=1}^{k}\mathsf{E}[Q_i(n)] = k\mathsf{E}[Q_i(n)]$, the mean service provided to each job by each of the $l$ servers will be $\kappa \mathsf{E}[Q_i(n)]$. The utilization of each server is then $\varrho = \lambda \kappa \mathsf{E}[Q_i(n)]$. Since $\lambda < 1/\mathsf{E}[\Delta(n)]$ for stability, the stability region, i.e., the maximum stable utilization for the tiny tasks model, is
\begin{align}
\varrho < \frac{\kappa \mathsf{E}[Q_i(n)]}{\mathsf{E}[\Delta(n)]} = \frac{1}{1 + \frac{1}{\kappa}\sum_{i=2}^l \frac{1}{i} }, & &\text{(tiny tasks)}
\label{eq:stabilitytinytasks}
\end{align}
where we inserted $\mathsf{E}[\Delta(n)]$ from Lem.~\ref{lem:tinytasksplitmerge} and $\mathsf{E}[Q_i(n)] = 1/\mu$.

For comparison, consider the equivalent big task split-merge model where the number of tasks $k$ equals the number of servers $l$ and $Q_i(n)\sim\text{Erlang}(\kappa,\mu)$.  From~\eqref{eq:splitmergeserver} the service process of the big task model is determined by the maximal task, $\Delta(n) = \max_{i \in [1,l]} \{Q_i(n)\}$. Since $\Delta(n)$ is non-negative, we can derive the expected value by integration of the complementary cumulative distribution function (CCDF) as
\begin{equation}
\begin{split}
\mathsf{E}[\Delta(n)] &= \int_0^{\infty} 1-\mathsf{P}[\Delta(n) \le x] dx \\
&= \int_0^{\infty} 1-(\mathsf{P}[Q_i(n) \le x])^l dx ,
\end{split}
\label{eq:splitmergeerlangjob}
\end{equation}
where we used that $\mathsf{P}[\max_{i \in [1,l]} \{Q_i(n)\} \le x] = \mathsf{P}[Q_1(n) \le x \cap Q_2(n) \le x \cap \dots \cap Q_l(n) \le x] = (\mathsf{P}[Q_i(n) \le x])^l$ for iid random variables $Q_i(n)$. Finally, we insert the Erlang-$\kappa$ CDF 
\begin{equation}
\mathsf{P}[Q_i(n) \le x] = 1 - e^{-\mu x} \sum_{i=0}^{\kappa-1} (\mu x)^i/i!
\label{eq:erlangcdf}
\end{equation}
and solve~\eqref{eq:splitmergeerlangjob} numerically. Again, $\lambda\mathsf{E}[\Delta(n)] < 1$ implies stability, and with $\varrho = \lambda \mathsf{E}[Q_i(n)]$, where $\mathsf{E}[Q_i(n)] = \kappa/\mu$ is the expected service time of the big tasks, the stability region for the big-tasks model follows as
\begin{align}
\varrho < \frac{\mathsf{E}[Q_i(n)]}{\mathsf{E}[\Delta(n)]} = \frac{\kappa}{\mu\mathsf{E}[\Delta(n)]} & &\text{(big tasks)}
\label{eq:stabilitybigtasks}
\end{align}
where $\mathsf{E}[\Delta(n)]$ is given by~\eqref{eq:splitmergeerlangjob}.

The stability region of the split-merge model with both big tasks and tiny tasks for $\kappa=20$ is shown in Fig.~\ref{fig:splitmergestabilitytinytasks}.  The model with tiny tasks shows a clear improvement of the stability region.

%
%------------------------------------------------------------------------
%------------------------------------------------------------------------
%
\subsection{Sojourn time bounds} 
Fig.~\ref{fig:sojourntimecomparisonsplitmergetinytasks} compares sojourn time bounds of the big tasks and tiny tasks models for equivalent parameters.  In the case of tiny tasks, the sojourn time bound is derived by substitution of parameter $\rho_S(\theta)$ from Lem.~\ref{lem:tinytasksplitmerge} into Th.~\ref{th:sojourntimesingleserver}. In the case of big tasks, we first have to derive the envelope rate $\rho_S(\theta)$ of the service process $S(m,n)$ defined in~\eqref{eq:splitmergeserver} for iid tasks with $Q_i(n)\sim\text{Erlang}(\kappa,\mu)$. We derive the MGF of $S(n)$ by integration of the CCDF as
\begin{equation*}
\mathsf{E}[e^{\theta S(n)}] = \int_0^{\infty} 1-\mathsf{P}[e^{\theta S(n)} \le x] dx .
\end{equation*}
Since for $\theta > 0$ it holds that $e^{\theta S(n)} \ge 1$ we have
\begin{equation*}
\mathsf{E}[e^{\theta S(n)}]  = 1 + \int_1^{\infty} 1-\mathsf{P}\left[S(n) \le \frac{\ln(x)}{\theta}\right] dx .
\end{equation*}
By definition of $S(n) = \max_{i \in [1,l]} \{ Q_i(n) \}$ it follows that $\mathsf{P}[S(n) \le x] = (\mathsf{P} [Q_i(n) \le x])^l$ so that
\begin{equation*}
\mathsf{E}[e^{\theta S(n)}]  = 1 + \int_1^{\infty} 1-\left(\mathsf{P}\left[Q_i(n) \le \frac{\ln(x)}{\theta}\right]\right)^l dx .
\end{equation*}
We insert the Erlang-$\kappa$ CDF~\eqref{eq:erlangcdf} and solve the integral numerically. The envelope rate follows as $\rho_S(\theta) = \ln (\mathsf{E}[e^{\theta S(n)}]) / \theta$ and the sojourn time bound is derived by use of Th.~\ref{th:sojourntimesingleserver}.

The sojourn time bounds in Fig.~\ref{fig:sojourntimecomparisonsplitmergetinytasks} are shown for iid inter-arrival times $A(n,n+1)\sim\text{Exp}(\lambda)$.  Three different $\lambda$ are used, corresponding to utilizations of $0.5$, $0.6$, and $0.7$.  The use of tiny tasks improves the sojourn time bounds significantly. The improvement is larger under higher utilizations, where the big tasks split-merge model becomes unstable for even relatively small numbers of servers $l$.

A more in-depth examination of the relationship between stability, performance, and idle times in the big tasks vs. tiny tasks cases is given in~\cite{fidler:tinytasks}.

%
%------------------------------------------------------------------------
%------------------------------------------------------------------------
%------------------------------------------------------------------------
%
\section{Single-queue fork-join with tiny tasks}
\label{sec:ideal-splitmerge}
\label{sec:sqfj-tiny-tasks-analysis}
We consider a single-queue fork-join model with tiny tasks.  The model is similar to the split-merge model with tiny tasks depicted in Fig.~\ref{fig:tinytaskssystem}, with one difference: there is no synchronization constraint at the start of a job.  I.e. a new job can start service as soon as a worker becomes idle and there are no unserviced tasks from the previous job.  As a consequence, workers will not idle at the end of a job if there are other jobs waiting.  Furthermore, jobs can overtake each other and finish service out of sequence.  For analytical tractability we study a model where the jobs depart in sequence, $D(n) \le D(n+1)$.  That is, jobs that finish service must wait in a queue until their predecessors have departed.  This added constraint does not affect the waiting time, $W(n)$, and means that the sojourn time bounds produced for this model will be strictly larger than those of the plain single-queue fork-join with tiny tasks model.

\begin{theorem}[Tiny tasks fork-join model]
\label{th:tinytaskforkjoin}
Given a fork-join model with $l$ servers and $k \ge l$ iid exponential tiny tasks with parameter $\mu$ and iid inter-arrival times with envelope rate $\rho_A(-\theta)$. For any $\theta \in (0,\mu)$ that satisfies $k\rho_Z(\theta) \le \rho_A(-\theta)$, the waiting time of task $i \in [1,k]$ of job $n \ge 1$ is bounded by
\begin{equation*}
\mathsf{P} \left[ W_i(n) \ge \tau \right] \le e^{\theta (i-1) \rho_Z(\theta)} e^{-\theta\tau} ,
\end{equation*}
and the sojourn time of job $n \ge 1$ by
\begin{equation*}
\mathsf{P} \left[ T(n) \ge \tau \right] \le e^{\theta ((k-1)\rho_Z(\theta)+\rho_X)} e^{-\theta\tau} .
\end{equation*}
The parameters $\rho_X(\theta)$ and $\rho_Z(\theta)$ are given in Lem.~\ref{lem:tinytasksplitmerge}.
\end{theorem}
As a special case for $k=l=1$, Th.~\ref{th:tinytaskforkjoin} recovers the single server case Th.~\ref{th:sojourntimesingleserver} for exponential jobs with envelope rate~\eqref{eq:rhoexposervice}. Also, for $k=l$, Th.~\ref{th:tinytaskforkjoin} recovers the waiting time bound for the single-queue fork-join model (without tiny tasks)~\cite[Th. 4]{fidler:multiserver}. For the sojourn time bound~\cite{fidler:multiserver} uses a slightly different derivation technique that can provide tighter bounds mostly at low utilizations.

\begin{figure}
\centering
\includegraphics[width=0.950\linewidth]{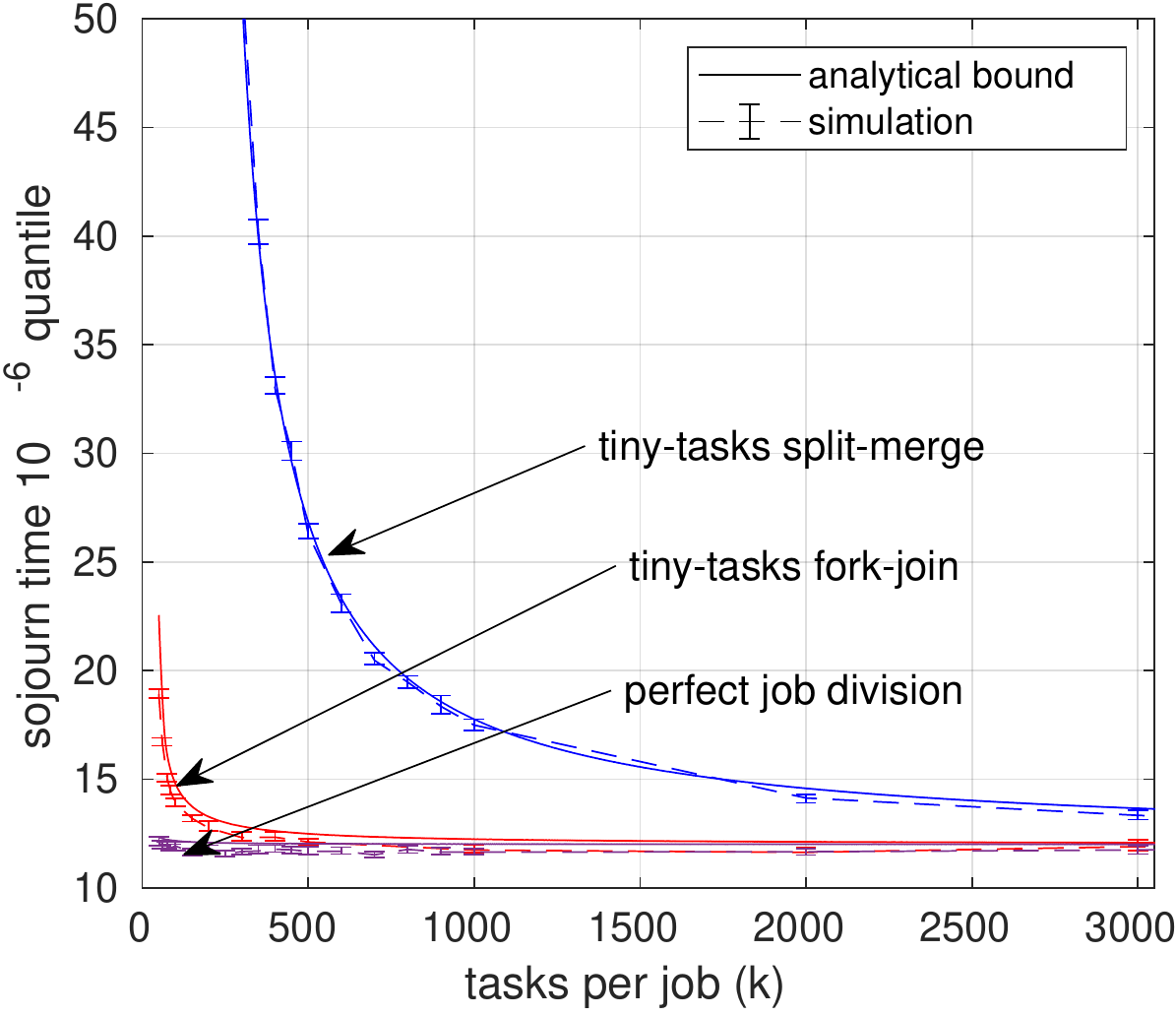}
\label{fig:sojourntimeconvergence2}
%}
\caption{Comparison of the sojourn time bounds of the single-queue fork-join and split-merge models with $l=50$ servers and $k$ tiny tasks. As a reference, the sojourn time bounds of a system with ideal partition, where a job is partitioned into $l$ equisized tasks, is shown. Jobs have exponential inter-arrival times with parameter $\lambda=0.5 \text{s}^{-1}$ and are composed of $k$ exponential tiny tasks with parameter $\mu = \frac{k}{l}$. The bounds are exceeded with probability at most $\varepsilon = 10^{-6}$.}
\label{fig:sojourntimeconvergence}
\end{figure}

The proof of Theorem~\ref{th:tinytaskforkjoin} is more involved, and can be found in~\cite{fidler:tinytasks}.  It is important to note that the random variables, $X$ and $Z$, represent the same things as in Lem.~\ref{lem:tinytasksplitmerge} and have the same MGFs.  This will be important when we incorporate overhead into the model in Sec.~\ref{sec:overhead-analyitical-models}.

Fig.~\ref{fig:sojourntimeconvergence} compares sojourn time bounds obtained for the single-queue fork-join and split-merge models with $l=50$ servers and a varying number $k$ of tiny tasks per job to the equivalent system with the ideal partition of jobs into $l$ equisized tasks.  The bounds in the figure are evaluated with violation probability $\varepsilon = 10^{-6}$.  As in Fig.~\ref{fig:sojourntime_sim_vs_spark} we increase the number of tasks per job, $k$,  and decrease the task service time proportionally, so that the mean job workload $\mathsf{E}[L(n)]$ remains constant.
%Specifically we use iid inter-arrival times $A(n,n+1)\sim\text{Exp}(\lambda=0.5)$, and iid task service times $Q_i(n)\sim\text{Exp}(\frac{k}{l})$ for $l=50$.  Therefore the jobs' workloads have a $L(n)\sim\text{Erlang}(k,\frac{k}{l})$ distribution, and constant mean $\mathsf{E}[L(n)]=l$ as we increase $k$.  The utilization follows as $\varrho = \lambda = 0.5$. 
In case of the ideal partition, we substitute $\mu=\frac{k}{l}$ into~\eqref{eq:rhoidealdivision} to get the corresponding envelope rate, $\rho_Q(\theta)$ that can be inserted into Th.~\ref{th:sojourntimesingleserver}.
%
%for an $\text{Erlang}(k,\mu)$ process the corresponding envelope rate would be
%%
%\begin{equation*}
%\rho_Q(\theta) = \frac{k}{\theta} \ln \left( \frac{\mu}{\mu-\theta} \right)
%\end{equation*}
%
%for $\theta \in (0,l\mu)$.  Substituting $\mu=l\frac{k}{l}=k$ we obtain $\rho_Q(\theta)$ that can be inserted into Th.~\ref{th:sojourntimesingleserver}.

As $k$ increases, the sojourn time bound of the fork-join model with tiny tasks quickly approaches that of the ideal partition.  For the tiny tasks split-merge model, for small $k$, the divergence between the models is quite large, a consequence of the restricted stability region of the split-merge model.  For large $k$, both models approach the ideal partition.

A more detailed look at the sojourn time bounds relative to the optimal, and how the utilization level affects the rate of convergence, is given in~\cite{fidler:tinytasks}.

%
%------------------------------------------------------------------------
%------------------------------------------------------------------------
%------------------------------------------------------------------------
%

\section{Incorporating overhead into the tiny-tasks analytical models}
\label{sec:overhead-analyitical-models}

Both Lemma~\ref{lem:tinytasksplitmerge} and Theorem~\ref{th:tinytaskforkjoin} define two random variables: $X(n)$ is the time from the start of task $k$ of job $n$ until the job departs, and $Z_i(n)$ is the time from the start of task $i>l$ of job $n$ until the next server becomes available.  As long as the task service time distributions are exponential, the memorylessness of the exponential distribution makes it possible to derive expressions and MGFs for $X(n)$ and $Z_i(n)$.  Without the assumption of memorylessness, the time remaining for each task-in-progress would depend on how long it had already been running, and these random variables would become extremely complicated if not impossible to solve for.

We have, however, created and simulated a detailed model of the overhead in a Spark system based on experimental measurements.  It is worth considering how this model can be incorporated into the analytical models to provide some approximation of system performance with overhead.  This is important for approximating the optimal range of $k$ and $l$ under differing overhead conditions.

Section~\ref{sec:overhead-distribution} identified two main classes of overhead: task-service overhead which effectively increases the service times of the tasks (it blocks subsequent tasks from starting), and pre-departure overhead that delays the departure of the job, but does not affect the processing of subsequent tasks within a job.  In the simulations we modeled the task overhead as having a constant and an exponential component~\eqref{eq:task-overhead-exponential}.  In the analytical model, a problem arises when we take the MGF of~\eqref{eq:task-overhead-exponential}, because the very small exponential limits us to $\theta <\mu^{ts}_{task}$.  For our purpose here, we model the entire task-service overhead using its mean
\begin{equation}
    \mathsf{E}\left( O_i(n)\right) = c^{ts}_{task} + \frac{1}{\mu^{ts}_{task}}
\end{equation}
This means that we neglect the outlier task overhead values noted in section~\ref{sec:overhead-distribution}.
%An alternative could be modeling the overhead using a Rayleigh distribution, which would have a reasonable distribution, and also a workable MGF, though it is atypical to use a Rayleigh distribution to model waiting times.

The pre-departure overhead was modeled in section~\ref{sec:overhead-distribution} using both a per-task and per-job constant~\eqref{eq:predeparture-overhead}.  In the fork-join case this is simple to model, since it just adds to each job's sojourn time, and does not affect the waiting time.  In the split-merge case, however, pre-departure overhead does not block subsequent tasks within a job, bit it does block subsequent jobs from starting.  It therefore does affect waiting times and needs to be incorporated into the analytical model.

%
%------------------------------------------------------------------------
%
\subsection{Overhead in the fork-join model}
\label{sec:overhead-fj-model}

We need to calculate how constant task overhead affects $X(n)$ and $Z_i(n)$ and their MGFs.  We refer to these modified random variables as $X^o(n)$ and $Z_i^o(n)$.  In the fork-join case, since the pre-departure overhead is non-blocking, $X^o(n)$ must be taken to be the time from the start of task $k$ of job $n$ until the job is ready to depart, absent the pre-departure overhead.  Therefore we have
\begin{equation}
    X^o(n) = X(n) + c^{ts}_{task} + \frac{1}{\mu^{ts}_{task}}
\end{equation}

A constant, $c$, has MGF $e^{c\theta}$, $\theta>0$.  Because the MGF of a sum of independent random variables is the product of MGFs, we can easily compute the MGF of $X^o(n)$, and then with Lem.~\ref{lem:tinytasksplitmerge}
\begin{equation}
    \rho_{X^o}(\theta) = c^{ts}_{task} + \frac{1}{\mu^{ts}_{task}} + \frac{1}{\theta} \sum_{i=1}^{l} \ln\left( \frac{i\mu}{i\mu - \theta} \right).
        \label{eq:overhead-rho-x-fj}
\end{equation}

We incorporate task overhead into $Z_i(n)$ similarly, but with a key difference.  $Z_i(n)$ is the time from the start of task~$i$ of job~$n$ until the next worker becomes available, and therefore has the distribution of the minimum of $l$ exponentials, which is $\text{Exp}(l\mu)$.  We can add the constant overhead to $Z_i(n)$, but this has the effect of adding the full task overhead to each active task each time a new task is scheduled.  On average, each task would pay the task overhead $l$ times during its execution.  We make the more reasonable approximation that each active task pays a $1/l$ fraction of the task overhead each time a new task is scheduled.  That is, we make the approximation
\begin{equation}
    Z_i^o(n) = Z_i(n) + \left(c^{ts}_{task} + \frac{1}{\mu^{ts}_{task}}\right)/l.
\end{equation}
which gives
\begin{equation}
    \rho_{Z^o}(\theta) = \frac{c^{ts}_{task} + \frac{1}{\mu^{ts}_{task}}}{l} + \frac{1}{\theta} \ln\left( \frac{l\mu}{l\mu - \theta} \right).
    \label{eq:overhead-rho-z}
\end{equation}

This allows us to evaluate an approximate waiting time distribution with overhead using Th.~\ref{th:tinytaskforkjoin}.  In order to approximate the distribution of sojourn time with overhead, we still need to add in the pre-departure overhead.  If $\tau_{\epsilon}$ is the sojourn time quantile approximation obtained using Th.~\ref{th:tinytaskforkjoin}, then we add
\begin{equation}
    \tau_{\epsilon}^o = \tau_{\epsilon} + c_{job}^{pd} + k \cdot c_{task}^{pd}.
    \label{eq:fj-sojourn-predeparture-overhead}
\end{equation}

The resulting approximate 0.99 sojourn time quantiles for $l=50$, $\lambda=0.5$, and $\mu=k$, and varying number of tasks per job ($k$) are plotted in Fig.~\ref{fig:sojourntime_sim_vs_spark_fj} along with corresponding simulation and spark experimental data.  The increase in the sojourn time due to task overhead matches very well with the simulation and experimental data.

%
%------------------------------------------------------------------------
%
\subsection{Overhead in the split-merge model}
\label{sec:overhead-sm-model}

Approximating the waiting and sojourn time distributions with overhead using Lem.~\ref{lem:tinytasksplitmerge} is handled in much the same way, except that now the pre-departure overhead is blocking.  This does not change the approximation for $Z_i^o(n)$, but now we have
\begin{equation}
    X^o(n) = X(n) + c^{ts}_{task} + \frac{1}{\mu^{ts}_{task}} + c_{job}^{pd} + k \cdot c_{task}^{pd}
\end{equation}
and we no longer add the pre-departure overhead directly to the sojourn time as in equation~\eqref{eq:fj-sojourn-predeparture-overhead}.  This gives
\begin{equation}
    \rho_{X^o}(\theta) = c^{ts}_{task} + \frac{1}{\mu^{ts}_{task}} + c_{job}^{pd} + k \cdot c_{task}^{pd} + \frac{1}{\theta} \sum_{i=1}^{l} \ln\left( \frac{i\mu}{i\mu - \theta} \right)
\end{equation}
which, along with~\eqref{eq:overhead-rho-z} can be used directly with Lem.~\ref{lem:tinytasksplitmerge}.  The resulting approximated sojourn time quantiles are plotted in Fig.~\ref{fig:sojourntime_sim_vs_spark_sm}, and again, fit the simulation and experimental results very well.  Importantly, in both the split-merge and fork-join cases, the analytical approximations provide good estimates for the optimal number of tasks-per-job, balancing the benefits of task tinyfication with the cost of scheduling and processing overhead.

%
%------------------------------------------------------------------------
%------------------------------------------------------------------------
%------------------------------------------------------------------------
%

\section{Conclusions}
\label{sec:conclusions}

Using ``tiny tasks'' in parallel processing systems refers to the practice of splitting jobs into $k$ tasks, where $k$ is larger than the number of workers, $l$, servicing the tasks.  Using moderately tiny tasks in map-reduce systems has long been a practical performance enhancement employed by practitioners, but at some point the performance benefit of dividing jobs into more smaller tasks is outweighed by the various types of system overhead.

We performed extensive experiments on an Apache Spark system using carefully controlled task size distributions to quantify the performance benefits of using tiny tasks, and measure and model the types of overhead that interfere.  We developed a model for how this overhead scales with job size, and implemented it in a simulator.  We presented analytical models looking at the improved stability region of split-merge systems using tiny tasks, and derived analytical bounds on the waiting and sojourn times of both split-merge and fork-join systems with tiny tasks.  Finally we used these analytical bounds, along with our model of scheduler overhead, to produce an analytical approximation for the sojourn time quantiles of both split-merge and fork-join systems with tiny tasks.  These approximations fit very well to our experimental and simulations results.

This work is important in understanding the fundamental properties of parallel systems and how they are affected by task granularity.  Our analytical approximation model which includes scheduling overhead can also be used to optimize task granularity on real systems.

\balance
\bibliographystyle{IEEEtran}
\bibliography{IEEEabrv,tinytasks}
%
%------------------------------------------------------------------------
%
\end{document}